\documentclass[12pt,a4paper]{article}
\usepackage{graphicx}
\usepackage[cp1251]{inputenc}
\usepackage[russian]{babel}
\usepackage{amssymb, amsfonts, amsmath}

\begin{document}


\title{Classification of Petrov Homogeneous Spaces}

\author{V. V.  Obukhov}

\maketitle

%
%
%


\quad  

Tomsk  State  Pedagogical University, Institute of Scietific Research and
Development,, 60 Kievskaya St., Tomsk 634041, Russia; obukhov@tspu.edu.ru\\

Tomsk State University of Control Systems and Radio Electronics,Laboratory for Theoretical Cosmology, International Center of Gravity and Cosmos, 36, Lenin Avenue, Tomsk 634050, Russia

\quad 

%


\section{Introduction}
Space-time manifolds $V_4$ admitting groups of motions occupy a special place in the theory of gravitation, since the vector Killing fields defining generators of these groups serve to construct the conserved physical quantities connected with the properties of the gravitational field. Therefore, these fields play an important role in describing the geometry of curved spacetime and physical phenomena occurring within~it.

A large number of papers have been devoted to this role of the Killing vector fields (see, for~example,~\cite{1,2,3,4}). The~question of group motions in Riemannian spaces was first considered in the works of Bianchi ~\cite{5,6},  Fubini and Rimini ~\cite{7,8} back in the late 19th and early 20th~centuries.

These spaces are of particular interest in cosmology, which, as a modern science, in~fact, began to develop after the appearance of works \cite{9,10}, in~which  special cases of homogeneous spaces  were used for the first time. These spaces are still the foundation of cosmological models, including in alternative theories of gravitation (see~\cite{11,11a}), as~well as in the early stages of the Universe's~development.

The classification of space-time manifolds with groups of motions is based on the theory of continuous groups of transformations (see~\cite{12,13}).

A complete classification of four-dimensional pseudo-Riemannian spaces founded on the canonical real structures of the motion groups acting in the spaces was realized by A.Z. Petrov in~\cite{14, 15, 16}. Petrov constructed non-equivalent sets of operators for all groups of motions of four-dimensional pseudo-Riemannian spaces, integrated the Killing equations and found the metrics of all these spaces.  When classifying isotropic spaces, Petrov used the approaches described in the papers by Kruchkovich~\cite{17,19}.

 The obtained results of this classification have found application in the theory of gravitation, first of all, when considering the problem of exact solutions of the field gravitational equations (see \cite{16,20, 37d} ).
 In~more recent works, various aspects of using spaces with groups of motions were considered. Thus, the authors of~\cite{21,22,23} studied the intersections of homogeneous and Stackel spaces, as~well as physical processes in homogeneous spaces. In~\cite{24,25}, the problems of constructing integrals of motion for classical and quantum scalar charged particles moving in spaces with groups of motions and in admissible electromagnetic fields were considered. The~classification of the exact solutions of Maxwell equations in homogeneous spaces in the presence of admissible electromagnetic fields was carried out in~\cite{26,27, 27a}. A~special direction of application of homogeneous spaces is the construction of the theory of noncommutative integration of the equations of motion of charged quantum particles in external fields~\cite{28,29,30,31}.

 In this paper, the final stage of the Petrov classification is carried out. As it is known, the Killing vector fields specify infinitesimal transformations of the group of motions of space $V_4$. In~the case where the group of motions $G_3$ acts in a simply transitive way in the homogeneous space $V_4$, the~geometry of the non-isotropic hypersurface is determined by the geometry of the transitivity space $V_3$ of the group $G_3$. In~this case, the~metric tensor of the space $V_3$ can be given by a nonholonomic reper consisting of three independent vectors $\ell_{(a)}^\alpha$, which define the generators of the group $G_3$ of finite transformations in the space $V_3$. The~representation of the metric tensor of $V_4$ spaces by means of vector fields $\ell_{(a)}^\alpha$ has a great physical meaning and makes it possible to substantially simplify the equations of mathematical physics in such spaces. Therefore, the~Petrov classification should be complemented by the classification of vector fields $\ell_{(a)}^\alpha$ connected to Killing vector fields. For~homogeneous spaces, this problem has been largely solved (see, e.g., \cite{36a, 33b, 35c}.  A~complete solution of this problem is presented in the present paper, where I refine ~the Petrov classification for homogeneous spaces in which the group $G_3$, which belongs to type $VIII$ according to the Petrov classification, acts simply transitively. In~addition, this paper provides the~complete classification of vector fields $\ell_{(a)}^\alpha$ for spaces $V_4$ in which the group $G_3$ 
 ~on isotropic~hypersurfaces.

\section{Homogeneous Petrov~Spaces}

Consider a Riemannian space $V_4$ with signature $(-1, 1, 1,1)$, on~the hypersurface $V_3$ of which the motion groups $G_3(N)$ act simply transitively ($N$ corresponds to the number of the group $G_3$ in the Bianchi classification). In~the case of a null hypersurface of transitivity $V^{*}_3(N)$, the~space $V^{*}_4(N)$ will be called a null homogeneous Petrov space of type $N$. If~the hypersurface of transitivity $V_3(N)$ is non-zero, then $V_4(N)$ will be called a homogeneous non-isotropic Petrov space of type $N$.

The geometry of a homogeneous Petrov space is connected with the geometry of the three-dimensional space of transitivity $\tilde{V}_3(N)$ of the group $G_3(N)$.
For a non-null homogeneous Petrov space, this connection is direct, since in this case  a hypersurface of transitivity $V_3(N)$  is in fact the space $\tilde{V}_3(N)$. For~a null Petrov space, this connection is less obvious, but~also exists, since the Killing vector fields of the space $V^{*}_4(N)$ are generally a linear combination of the Killing vector fields of the space $\tilde{V}_3(N)$ with coefficients depending on the wave variable $u^0$.

 The contravariant components of the metric tensor in a non-null semi-geodesic coordinate system can be represented in the following form:
\begin{equation} \label{1}
g^{ij} =\left(\begin{array}{cccc} {g^{\alpha \beta } } & {\left(u^{i} \right)} & {} & {0} \\ {} & {} & {} & {0} \\ {} & {} & {} & {0} \\ {0} & {0} & {0} & {\varepsilon} \end{array}\right) \quad (\varepsilon \pm 1).
\end{equation}
The admissible transformations of the coordinates $u^{i}$, which do not violate the form of \eqref{1}, are of the following form:
\begin{equation} \label{2}
\tilde{u}^{\alpha} =\tilde{u}^{\alpha}(u^\beta), \quad \tilde{u}^{0} = u^0.
\end{equation}
It follows from the Killing equations that
\begin{equation}\label{3}
 \begin{array}{cc} {\xi _{\left(a\right),0}^{\alpha} = 0,} & {\xi ^{0}_{(a)} =0}. \end{array}
\end{equation}
The following index designations are used here and hereafter:
$$
 \begin{array}{ccc} {i,j,r,\ell \div 1,2,3,0.} & {\alpha ,\beta ,\gamma \div 1,2,3.} & {a,b,c\div 1,2,3.} \end{array}
$$
Zero indices number the fourth row and column.
Greek letters denote the coordinate indices of the semi-geodetic coordinate system $u^{i}$. The~small Latin letters a, b, and c~denote the indices of the nonholonomic coordinate system associated with the group $G_{3} (N)$.

The classification of null homogeneous Petrov spaces was carried out in the isotropic semi-geodesic coordinate system (see~\cite{14}, p. 158). In~this coordinate system, the contravariant components of the metric tensor $g^{ij} $ have the form
\begin{equation}\label{4}
g^{ij} =\left(\begin{array}{cccc} {g^{\alpha \beta }(u^{i}) } & {} & {} & {A(u^0,u^p)} \\ {} & {} & {} & {0} \\ {} & {} & {} & {0} \\ {A(u^0,u^p)} & {0} & {0} & {0} \end{array}\right)
\end{equation}
There and further, the~indices denoted by $ p, q $ change within $\div 1,2$.
In the present paper,  the canonical form of the isotropic semi-geodesic coordinate system, in~which
\begin{equation}\label{5}
A=1,
\end{equation}
is used:
\begin{equation} \label{6}
g^{ij} =\left(\begin{array}{cccc} {g^{\alpha \beta }(u^{i}) } & {} & {} & {1} \\ {} & {} & {} & {0} \\ {} & {} & {} & {0} \\ {1} & {0} & {0} & {0} \end{array}\right).
\end{equation}
It follows from the Killing equations that in the canonical coordinate system the components of the Killing vector fields of the group $G_3(N)$ satisfy the following conditions (see~\cite{16}):
\begin{equation} \label{7}
\xi _{\left(a\right),1}^{\alpha } =\xi _{\left(a\right)}^{0} =0.
\end{equation}
The admissible coordinate transformations $\left\{u^{i} \right\}$, which do not violate the conditions \eqref{6}, \eqref{7}, are of the following form:
\begin{equation} \label{8}
\begin{array}{cc} {\widetilde{u}^{0}=\varphi_0(u^0), \quad \widetilde{u}^{1} =u^{1}/\varphi_{0,0}+\varphi ^{1} \left(u^{0}, u^{p} \right),} & {\widetilde{u}^{p} =\varphi ^{p} \left(u^{0} ,u^{q}\right)} \end{array}
\end{equation}
The homogeneous space $V_3(N)$ has two equivalent definitions. According to the first one, $V_3(N)$ is the space of transitivity for the group $G_3(N)$, whose operators are constructed with  Killing vector fields $\xi_{(a)}^\alpha$:
\begin{equation}\label{9}
{\rm X}_a =\xi_{(a)}^\alpha {\rm p}_\alpha,
\end{equation}
According to the second definition, in the homogeneous space $V_3(N)$ the group $G_3(N)$  of non-infinitesimal shifts acts. The~group operators have the form
\begin{equation}\label{10}
{\rm Y}_a =\ell_{(a)}^\alpha {\rm p}_\alpha.
\end{equation}
 According to the first definition, two infinitely close points of space $V_3(N)$ can be combined using the motion group of space $G_3(N)$. According to the second definition, this can be done for any two points of the space. As~is known (see, for~example, \cite{36a})
  p. 483), the~metric tensor of the homogeneous space $V_3(N)$ is expressed through the vector fields $\ell^\alpha_{(a)}$ as follows:
\begin{equation}\label{11}
g^{\alpha\beta} = \ell^\alpha_{(a)}\ell^\beta_{(b)} \eta^{ab}.
\end{equation}
${\alpha,\beta,\gamma \div 1,2,3}; \eta^{ab}=const$  (in the case where  $\eta^{ab} $ is considered part of the metric tensor $g^{ij},  \eta^{ab}$  are functions of the variable $u^0$). The sets of operators ${\rm X}_a,  {\rm Y}_b$  obey the same structure equations:
\begin{equation}\label{12}
[{\rm X}_a,{\rm X}_b] = C^c_{ab}{\rm X}_c,
\end{equation}
\begin{equation}\label{13}
[{\rm Y}_a,{\rm Y}_b] = C^c_{ab}{\rm Y}_c.
\end{equation}
Hence, the~sets of vector fields $\{\xi_{(a)}^\alpha \}$ and $\{\ell_{(a)}^\alpha \}$ are equivalent to each other with respect to coordinate transformations.
Therefore, both fields can be simultaneously considered as  Killing vector fields and as nonholonomic reper vector fields (see \eqref{11}) in the same space $V_3(N)$  but in different coordinate systems connected by admissible transformations of the form \eqref{2} or \eqref{8}.
In one coordinate system, these fields are connected by relations (see \cite{36a}, p. 484):
\begin{equation} \label{14}
\ell _{\left(a\right),\beta }^{\alpha } =\xi _{\beta }^{\left(b\right)} \xi _{\left(b\right),\gamma }^{\alpha } \ell _{\left(a\right)}^{\gamma } \Rightarrow
\xi_{\left(a\right),\beta }^{\alpha } =\ell _{\beta }^{\left(b\right)} \ell _{\left(b\right),\gamma }^{\alpha } \xi_{\left(a\right)}^{\gamma },\end{equation}
where
\begin{equation}\label{15}
\xi_{\beta }^{(b)} \xi_{(b)}^\alpha =\ell_{\beta}^{(b)} \ell _{(b)}^\alpha = \delta_\beta^\alpha.
\end{equation}
 These considerations apply to non-null homogeneous Petrov spaces. In~the case of null Petrov spaces, they cannot be used directly because the covariant metric tensor of the space $V_3(N)$ is not the metric tensor of the isotropic hypersurface of the space $V^{*}_4$. Therefore, the~proof of the \eqref{4} relations based on the above second definition of homogeneous space
  for null spaces is inapplicable.
 Nevertheless, the~representation of \eqref{11}, where the vector fields  $\ell _{\left(a\right),\beta }^{\alpha }$  satisfy the system of Equation \eqref{14}, holds for null homogeneous Petrov spaces as well. It can be shown that if
\begin{equation}\label{16}
det|\zeta_a^\alpha| \ne 0   \Rightarrow \zeta_a^\alpha\zeta_a^\beta =\delta^\beta_\alpha,
\end{equation}
vector fields satisfy a system of equations of the form \eqref{14}, then $\zeta_a^\alpha$ are Killing vector fields for a space with metric tensor \eqref{11}.
Indeed, substituting the expressions
\begin{equation}\label{17}
\zeta_{\left(a\right),\beta }^{\alpha } =\ell _{\beta }^{\left(b\right)} \ell _{\left(b\right),\gamma }^{\alpha }\zeta_{\left(a\right)}^{\gamma }
\end{equation}
and \eqref{11} into the Killing equations, we obtain the identity. As~Equation \eqref{17} has three independent solutions, the~vector fields $\zeta_{(a),\beta }^{\alpha} $ (also as $\ell_a^\alpha $ ) form the group $G_3$.

The classification has been carried out by Petrov in three~stages.

1. Selection of a semi-geodesic coordinate system.

 2. Solving structural Equation \eqref{4}.
In the second stage,  all sets of Killing vector fields that are non-equivalent with respect to admissible transformations of variables have been listed. In~this way, the~choice of semi-geodesic coordinate systems in which the solutions of the systems of structural equations can be represented in elementary functions is also~realized.

3. Solving the Killing equations.
In the third step, using the solutions found in the second step, the~Killing equations have been integrated:
\begin{equation}\label{18}
g^{\alpha\beta}_{,\gamma}\xi^\gamma_a = g^{\alpha\gamma} \xi^\beta_{a,\gamma} + g^{\beta\gamma} \xi^\alpha_{a,\gamma},
\end{equation}
 and the metric tensors of all the appropriate homogeneous Petrov spaces have been~found.

\quad

Representation of the metric tensor of homogeneous space in the form \eqref{11} has a deep physical meaning, and it is demanded when solving problems of mathematical physics in the theory of gravitation. Therefore, in~this paper, the classification of homogeneous Petrov spaces has been supplemented by the fourth~stage.

4. Solving the systems of Equation \eqref{14}.
In the fourth stage, all non-equivalent solutions of the systems of Equation \eqref{14} have been found. When the components of the Killing vector fields do not depend on the wave variable $u^0$, the~components of the metric $g^{\alpha\beta}$ are found directly from the \eqref{16} relations without directly using the Killing equations. Let us consider the peculiarities of the implementation of this step in the presence of such~dependence.

Each of the groups $G_3(II), G_3(III), G_3(V)$  has several independent sets of group operators. One of these sets contains a linear combination of group operators of the following form (see~\cite{16}):
$$
\tilde{{\rm X}}_{a^{'}} = {\rm X}_{a^{'}} + \ell_0(u^0) {\rm X}_1, \quad {\rm X}_1 = {\rm p}_2, \quad {\rm X}_{a,0} = (\xi^\alpha_{(a)}{\rm p}_\alpha)_{,0} =0, \quad [{\rm X}_a,{\rm X}_b] = C^c_{ab} {\rm X}_c.
$$
Using the solutions of the system of Equation \eqref{15}, one can find the contravariant components \eqref{7} of the metric tensor $g^{\alpha\beta}$ of the space $V^{*}_4$ admitting Killing vector fields $\xi^\alpha_{(a)}$. Then the contravariant components of the metric tensor $\tilde{g}^{\alpha\beta}$ can be represented as
\begin{equation}\label{19} \tilde{g}^{\alpha\beta} = g^{\alpha\beta} + \dot{\ell}_0\Delta g^{\alpha\beta}.
\end{equation} The function $\Delta g^{\alpha\beta}$ is found from the Killing equations: $$ \tilde{g}^{\alpha\beta}_{,\gamma}\tilde{\xi}^\gamma_a = \tilde{g}^{\alpha\gamma} \tilde{\xi}^\beta_{a,\gamma} + \tilde{g}^{\beta\gamma} \tilde{\xi}^\alpha_{a,\gamma}. $$
Here and hereafter,
$$
\dot{\ell}_0 = \ell_{0,0}.
$$

\section{The Diagonalization of Group~Vectors }

When making the classification  of homogeneous Petrov spaces, one can restrict oneself to the null case only, since the group of admissible coordinate transformations \eqref{8} is a subgroup of the group of admissible coordinate transformations \eqref{2}. Therefore, all sets of the group operators $G_3(N)$, which act in the space $V^{*}_4(N)$, form equivalence classes with respect to the transformation group \eqref{2} for the set of operators of the same group, which acts in the space $V_4(N)$. One feature  must be borne in mind when constructing the metric tensor $g^{\alpha \beta }$ according to the formula \eqref{11}.  In~the non-null space $V_4(N)$, the number of independent functions $\eta^{ab}(u^0)$ cannot be reduced by admissible transformations of \mbox{form \eqref{2}.}

The first step of the classification  is to choose a coordinate system, in~which  one of the  Killing fields has to be represented in the form
\begin{equation} \label{1a}
\xi _{\left(1\right)}^{\alpha } =\delta _{1}^{\alpha } {\rm A_1} (u^{0},u^{p}).
\end{equation}
Using admissible coordinate transformations \eqref{2}, it can always be made  for any vector. Moreover, it is possible to make the function ${\rm A} $ equal to unity. However, using admissible transformations \eqref{8} this can be carried out only for the vector that is already in this shape.
 Indeed, under~the transformation \eqref{8}, the operator ${\rm X} _{a} $ is transformed as follows:
\begin{equation} \label{2a}
\widetilde{{\rm X} }_{a} =\left({\rm A} _{a} +{\rm B} _{a} \Phi _{2}^{1} +R_{a} \Phi _{,3}^{1} \right)\widetilde{{\rm p} }_{1} +\left({\rm B} _{a} \Phi _{2}^{2} +R_{a} \Phi _{,2}^{3} \right)\widetilde{{\rm p} }_{2} +\left({\rm B} _{a} \Phi _{3}^{2} +R_{a} \Phi _{,3}^{3} \right)\widetilde{{\rm p} }_{3} .
\end{equation}
The following notations are used here and hereafter:
$${\rm X}_{a} ={\rm A}_{a}{\rm p}_1 +{\rm B}_{a}{\rm p}_2 +{\rm R}_{a}{\rm p}_3 \quad \tilde{u}^1 = u^1 + \Phi^1, \quad \tilde{u}^p = \Phi^p. $$
The letters  $\Phi^{\alpha }, {\rm A} _{a}, {\rm B} _{a},$ and $ {\rm R}_{a} $ denote functions depending on the variables $u^{0}$ and $ u^{p} $. Functions $\varphi^{a}, {a}_{a}, b_{a}, r_{a} $ depend on $u^{0} $ and one of the variables $u^{p}$.
Since the Jacobian of the transformation \eqref{8} does not equal to zero, it is possible to represent $\tilde{X}_{1} $ in the form \eqref{1a} only if ${\rm B} _{1} ={\rm R} _{1} =0$. In~this case, any of the remaining operators, for~example, ${\rm X} _{2} $, can be diagonalized using admissible transformations \eqref{8}. Indeed, without~restriction of generality, let us require that in a new coordinate system ${\{\tilde{u}\}}$ the operator ${\rm X} _{2}$  has the form
$$
{\rm X}_{2} ={\rm p }_{2}.
$$
For this purpose, the~following 
condition must be fulfilled:
\begin{equation} \label{3a}
\left\{\begin{array}{c} {{\rm A} _{2} + {\rm B} _{2} \Phi _{,2}^{1} +{\rm R}_{2} \Phi _{,3}^{1} =0} \\\ {{\rm B} _{2} \Phi _{,2}^{2} +{\rm R}_{2} \Phi _{,2}^{3} =1} \\\ {{\rm B} _{2} \Phi _{,3}^{2} +{\rm R}_{2} \Phi _{,3}^{3} =0} \end{array}\right.
\end{equation}
The system \eqref{3a} reduces to the form of the Cauchy--Kovalevskaya system and is consistent.
Thus, operator ${\rm X}_2$ can be presented in the following form:
\begin{equation} \label{4a}
{\rm X} _{2} ={\rm p}_2.
\end{equation}
The first option is~received.

\noindent
1. First version:

$${\rm X}_{1}=a_1(u^0, u^3){\rm p }_{1}, \quad {\rm X}_{2} ={\rm p }_{2}, \quad {\rm X}_{3} ={\rm A}_{3}{\rm p}_1 +{\rm B}_{3}{\rm p}_2 +{\rm R}_{3}{\rm p}_3.$$
Obviously, this version is valid only for solvable groups. Hereinafter, operators  ${\rm X}_{1}, {\rm X}_{2}$  are chosen as operators of the abelian subgroups of the groups
 $G_3(I) - G_3(VII)$. Therefore, from~${\rm A}_{,2} = 0$ it follows that   ${\rm A}_1 = a_1(u^0,u^3).$

\noindent
If  $|{\rm B}_{1}|+ |{\rm R}_{1}|\ne 0 $, from~\eqref{3a} it follows that ${\rm X}_{1}$ can be presented in the form
\begin{equation}\label{5a}
{\rm X} _{1} ={\rm p}_{2}.
\end{equation}
The admissible coordinate transformations that do not violate the condition \eqref{5a} are as follows:
\begin{equation} \label{6a}
\widetilde{u}^{0}=\varphi_0, \quad \widetilde{u}^{1} =\frac{u^{1}}{\varphi_{0,0}} +\varphi^{1} \left(u^{0},u^{3} \right), \quad \widetilde{u}^{p} =\varphi^{p} \left(u^{0},u^{3} \right).
\end{equation}
Operator ${\rm X} _{2} $ commutes with operator ${\rm X} _{1} $. Therefore,  after the admissible transformations \eqref{6a} it takes the following form:
\begin{equation} \label{7a}
{\rm X} _{2}=\left(a_{2} +{\rm r}_{2} \varphi _{,3}^{3} \right){\rm p} _{1} +\left(b_{2} +{\rm r}_{2} \varphi _{,3}^{2} \right){\rm p} _{2} +{\rm r}_{2} \varphi _{,3}^{3} {\rm p} _{3}.
\end{equation}
Obviously, only the following options are~possible:
\begin{enumerate}
\item $r_{2} \ne 0\Rightarrow$ \quad ${\rm X} _{2}={\rm p} _{3} $;

\item $r_{2}=0\Rightarrow {\rm X} _{2}=a_{2}{\rm p} _{1} +b_{2} {\rm p} _{2}\quad (a_{2,2}=b_{2,2}=0) $.
\end{enumerate}
In  the last case, one can use the transformations \eqref{6a} to simplify operator  ${\rm X}_{3}$.
Thus, these options together with versions for unsolvable groups have to be considered separately when making the classification.

\noindent
2. Second version:
\begin{equation}\label{8a}
{\rm X}_{1}={\rm p }_{2}, \quad {\rm X}_{2} ={\rm p }_{3}, \quad {\rm X}_{3} ={\rm A}_{3}{\rm p}_1 +{\rm B}_{3}{\rm p}_2 +{\rm R}_{3}{\rm p}_3.
\end{equation}

\noindent
3. Third version:
\begin{equation}\label{8a}
{\rm X}_{1}={\rm p }_{2}, \quad {\rm X} _{2}=a_{2}{\rm p} _{1} +b_{2} {\rm p} _{2}, \quad {\rm X}_{3} ={\rm A}_{3}{\rm p}_1 +{\rm B}_{3}{\rm p}_2 +{\rm R}_{3}{\rm p}_3.
\end{equation}
When solving structural equations, admissible transformations \eqref{6a} can be used to simplify the form of the operator ${\rm X}_{3}.$

For insoluble groups $G_3(VIII), G_3(IX)$ there is following version:

\noindent
4. ourth version:
\begin{equation}\label{9a}
{\rm X}_{1}={\rm p }_{2}, \quad {\rm X}_{p} ={\rm A}_{p}{\rm p}_1 +{\rm B}_{p}{\rm p}_2 +{\rm R}_{p}{\rm p}_3.
\end{equation}
In this case, the admissible transformations have the form \eqref{6a}. They can be used to simplify the form for one of the operators ${\rm X}_{p}.$

This paper is constructed as follows.

\begin{itemize}
\item[1.] Since the method used in this paper for classifying non-equivalent sets of motion group operators differs somewhat from the method used in~\cite{16} (in particular, admissible coordinate transformations of the form \eqref{8a} are used), some details of the computations in solving the structural equations are given below. Some of the results obtained have a simpler form than those given in Petrov's book. In~addition, I
 ~managed to complete the Petrov classification with two new  homogeneous non-isotropic  spaces of type $V_4(VII)$.

\item[2.] For~each non-equivalent set of group operators constructed using Killing vector fields, the classification of the reper vectors $\ell_{(a)}^\alpha$ is~given.

\item[3.] Using formulas \eqref{11} and  \eqref{19}, the~components of the contravariant metric tensor are constructed for each non-equivalent set of group~operators.

\item[4.] In~the final section, all non-equivalent sets of the group operators ${\rm X} _{2}$ and~${\rm Y} _{2}$ are~listed.
\end{itemize}

\section{Solvable~Groups}

\subsection{${Group \quad  G_{3}(I)}$}

Since three mutually commuting vector fields can always be diagonalized, even by using coordinate transformations of the form \eqref{8} \cite{13}, the~metric tensor of an isotropic space of  type $I$ according to the Bianchi classification can be transformed to the following form (
$\xi _a^{\alpha } =\delta _{a}^{\alpha } $): 
$$
g^{ij} =\left(\begin{array}{cccc} {\eta ^{\alpha \beta }  } & {\left(u^{0} \right)} & {} & {1} \\ {} & {} & {} & {0} \\ {} & {} & {} & {0} \\ {1} & {0} & {0} & {0} \end{array}\right),
$$
where $\eta^{\alpha\beta} = a_{\alpha\beta}(u^{0})$. It contains 6 independent functions of $u^{0} $. The~number of these functions can be reduced by an admissible transformation of variables:
\begin{equation} \label{1b}
\widetilde{u}^{\alpha} =u^{\alpha} +\varphi^{\alpha} \left(u^{0} \right)
\end{equation}
As a result, the functions $a_{1\alpha}$ can be inverted to zero and the metric tensor has the form
\begin{equation} \label{2b}
g^{ij} =\left(\begin{array}{cccc} {0 } & {0} & {0} & {1} \\ {0} & {a_{22}} & {a_{23}} & {0} \\ {0} & {a_{23}} & {a_{33}} & {0} \\ {1} & {0} & {0} & {0} \end{array}\right).
\end{equation}

\subsection{${Group \quad G_{3}(II)}$}

The structural equations have the following  form:
\begin{equation}\label{3b}
\left[{\rm X} _{1} {\rm X} _{2} \right]=0, \quad \left[{\rm X} _{2},{\rm X} _{3} \right]={\rm X} _{1}, \quad \left[{\rm X} _{1} {\rm X} _{3} \right]=0.
\end{equation}
 Let us consider all versions presented~above.

\noindent
First version. The~operators ${\rm X} _{a}$ are of the following form:
\begin{equation} \label{4b}
{\rm X} _{1} =a_{1} {\rm p}_{1}, \quad {\rm X} _{2} ={\rm p}_{2}, \quad {\rm X}_{3} =A_{3} {\rm p}_{1} + B_{3}{\rm p}_{2} +R_{3}{\rm p}_{3}
\end{equation}
From the third equation of system \eqref{3b}, it follows that $a_1 = 1$.
Using admissible transformation \eqref{6a}, one can find solution of the second equation from the system \eqref{3b} in the form
\begin{equation} \label{5b}
{\rm X} _{1} ={\rm p}_{1}, \quad {\rm X} _{2} ={\rm p}_{2}, \quad {\rm X} _{3} = u^{2} {\rm p}_{1} +{\rm p}_{3}.
\end{equation}
Let us construct the matrices:
\begin{equation}\label{6b}
\xi^\alpha_{(a)} = \begin{pmatrix} 1 & 0& 0 \\0 & 1 & 0 \\u^2 &0 & 1 \end{pmatrix}, \quad
\xi^{(a)}_\alpha = \begin{pmatrix} 1 & 0& 0 \\0 & 1 & 0 \\-u^2 & 0 & 1 \end{pmatrix}, \quad
\xi^{(b)}_\beta \xi^\alpha_{(b),\gamma} = \delta_\gamma^2\begin{pmatrix}0 & 0& 0 \\0 & 0 & 0\\1 & 0 & 1 \end{pmatrix}. \quad
\end{equation}
Then, the set of Equation \eqref{15} takes the form
\begin{equation}\label{7b}
\ell_{(a),1}^1 = \ell_{(a),\beta}^p =0, \quad \ell_{(a),3}^1 = \ell_{(a)}^2.
\end{equation}
From \eqref{7b}, it follows that
\begin{equation} \label{8b}
\ell _{\left(a\right)}^{\alpha } =\delta _{1}^{\alpha } \left(\delta _{a}^{1} +u^{3} \delta _{a}^{3} \right)+\delta _{2}^{\alpha } \delta _{a}^{2} +\delta _{3}^{\alpha } \delta _{a}^{3}
\end{equation}
From \eqref{8b}, one can obtain the matrix  $g^{\alpha \beta } =\ell _{\left(a\right)}^{\alpha } \ell _{\left(b\right)}^{\beta } \eta ^{ab} $:
\begin{equation} \label{9b}
g^{\alpha \beta } =\left(\begin{array}{ccc} {a_{11} +2a_{13} u^{3} +a_{33} u_{3}^{2} } & {a_{12} +u^{3} a_{23} } & {a_{13} +u^{3} a_{33} } \\\ {a_{12} +u^{3} a_{23} } & {a_{22} } & {a_{23} } \\\ {a_{13} +u^{3} a_{33} } & {a_{23} } & {a_{33} } \end{array}\right)
\end{equation}
The number of independent functions in the metric tensor \eqref{9b} can be reduced to three by admissible coordinate~transformations.

\noindent
Second version. The operators ${\rm X} _{a} $ are of the following form: $${\rm X} _{1} =p_{2} ,\quad {\rm X} _{2} =p_{3} ,\quad {\rm X} _{3} ={\rm A}_{3} p_{1} +{\rm B}p_{2} +{\rm R} p_{3}.
$$ From  structure Equation \eqref{3b}, it follows that
\begin{equation} \label{10b}
{\rm X} _{1} ={\rm p}_{2}, \quad {\rm X} _{2} ={\rm p}_{3}, \quad \tilde{{\rm X}} _{3} = {\rm X} _{3}+2{\ell_{0}} {\rm p}_{3}, \quad {\rm X} _{3}= {\rm p}_{1} + u^{3} {\rm p}_{2}
\end{equation}
One can find the vectors $\ell _{a}^{\alpha }$
using the following operators:
$${\rm X} _{1} ={\rm p}_{2}, \quad {\rm X} _{2} ={\rm p}_{3}, \quad {\rm X} _{3} ={\rm p}_{1} + u^{3} {\rm p}_{2}
$$
Let us construct the matrices:
\begin{equation}\label{11b}
\xi^\alpha_{(a)} = \begin{pmatrix} 1 & u^3 & 0 \\0 & 1 & 0 \\0 &0 & 1 \end{pmatrix}, \quad
\xi^{(a)}_\alpha = \begin{pmatrix} 1 & -u^3 & 0 \\0 & 1 & 0 \\0 & 0 & 1 \end{pmatrix}, \quad
\xi^{(b)}_\beta \xi^\alpha_{(b),\gamma} = \delta_\gamma^3\begin{pmatrix} 0 & 1& 0 \\0 & 0 & 0 \\0 & 0 & 0 \end{pmatrix}. \quad
\end{equation}
Then, the set of Equation \eqref{14} takes the following form:
\begin{equation}\label{12b}
\ell_{(a),p}^\alpha = \ell_{(a),\beta}^1 = \ell_{(a),\beta}^3 =0, \quad \ell_{(a),2}^1 = \ell_{(a)}^2.
\end{equation}
From \eqref{12b}, it follows that
\begin{equation} \label{13b}
\ell _{\left(a\right)}^{\alpha } =\delta _{1}^{\alpha } \delta _{a}^{1} + \delta _{2}^{\alpha } \left(\delta _{a}^{2} +u^{1} \delta _{a}^{3} \right)+\delta _{3}^{\alpha } \delta _{a}^{3}
\end{equation}
Let us find the matrix
$${g}^{\alpha \beta } =\ell _{\left(a\right)}^{\alpha } \eta ^{ab} \ell _{\left(b\right)}^{\beta }  =\begin{pmatrix}{a_{11} } & a_{12} +u^{1} a_{13} & a_{13} \\ {a_{12} + a_{13}u^1 } & {a_{22} +2u^{1} a_{23}u_1 +u_{1}^{2} a_{33} } & {a_{23} +u^{1} a_{33} } \\ {a_{13}} & {a_{23} +u^{1} a_{33} } & {a_{33} } \end{pmatrix}
$$

\noindent From the Killing equations, we obtain
$$G^{\alpha \beta } =\dot{\ell }_{0}u_{1}\left(2 (\delta _{1}^{\alpha } \delta _{3}^{\beta } +\delta _{3}^{\alpha } \delta _{1}^{\beta } )  +  u_{1}(\delta _{1}^{\alpha } \delta _{2}^{\beta } +\delta _{2}^{\alpha } \delta _{1}^{\beta }) \right)$$
Finally, we obtain the following:
\begin{equation} \label{14b}
\widetilde{g}^{\alpha\beta}= G^{\alpha \beta } + g^{\alpha \beta } =\begin{pmatrix}{a_{11} } & {a_{12} +u^{1} a_{13} + \dot{\ell}_0 {u^1}^2 } & {a_{13} +2\dot{\ell}_0 u^{1} }  \\ {a_{12} + a_{13}u^1 +\dot{\ell}_0 {u^1}^2  } & {a_{22} +u^{1} a_{23} +u_{1}^{2} a_{33} } & {a_{23} +u^{1} a_{33} } \\ {a_{13} +2\dot{\ell}_0 u^{1}} & {a_{23} +u^{1} a_{33} } & {a_{33} } \end{pmatrix}.\end{equation}
Using the admissible transformations of coordinates,  one can show that $g^{\alpha\beta}$ contains four independent functions of $u^{0} $.

\subsection{${Group \quad G_{3}(III)}$}

\noindent
Let us write the structural equations in the following  form:
\begin{equation} \label{1c}
\left[{\rm X} _{1} {\rm X} _{2} \right]=\left[{\rm X} _{1} {\rm X} _{3} \right]=0, \quad \left[{\rm X} _{2} {\rm X} _{3} \right]={\rm X} _{2}.
\end{equation}

\noindent
First version.
The operators ${\rm X} _{a} $ are of the following form:
$${\rm X} _{1} =a_{3} {\rm p} _{1}, \quad {\rm X} _{2} ={\rm p}_{2}, \quad {\rm X} _{3} ={\rm A} _{3} \rm p_{1} +{\rm B} _{3} \rm p _{2} +R_{3} \rm p_{3}. $$
From the equations of the system \eqref{1c}, it follows that

\noindent $a_{3} =1,   {\rm A}_{3}=0,  B_3 = u^2,  {\rm R}_{3} = 1.$
Thus, in~this case, the~set of group operators has the following  form:
\begin{equation} \label{2c}
{\rm X} _{1} =\rm p_{1}, \quad {\rm X} _{2} =\rm p_{2}, \quad{\rm X} _{3} =\rm p_{3}+u^{2}\rm p_{2}.
\end{equation}
Using the following matrices,
 $$\xi _{\left(a\right)}^{\alpha } =\left(\begin{array}{ccc} {1} & {0} & {0} \\ {0} & {1} & {0} \\ {0} & {u^{2} } & {1} \end{array}\right),\quad \xi _{\alpha }^{\left(a\right)} =\left(\begin{array}{ccc} {1} & {0} & {0} \\ {0} & {1} & {0} \\ {0} & {-u^{2} } & {1} \end{array}\right), \quad \xi^{(b)}_\beta \xi^\alpha_{(b),\gamma} =\delta _{\gamma }^{2} \left(\begin{array}{ccc} {0} & {0} & {0} \\ {0} & {0} & {0} \\ {0} & {1} & {0} \end{array}\right),$$
one can find the system of Equation \eqref{15} in the following  form:
\begin{equation} \label{3c}
\ell _{\left(a\right),\beta }^{\alpha } = \delta^\alpha_2 \delta^3_\beta \ell _{\left(a\right)}^{2}.
\end{equation}
The solution of  system \eqref{3c} is
\begin{equation} \label{4c}
\ell _{\left(a\right)}^{\alpha } =\delta _{1}^{\alpha } \delta _{a}^{1} +\delta _{2}^{\alpha } \delta _{a}^{2} \exp \left(-u_{3} \right)+\delta _{3}^{\alpha } \delta _{a}^{3}
\end{equation}
Using \eqref{4c}, one can obtain the matrix $g^{\alpha \beta } $:
\begin{equation} \label{5c}
g^{\alpha \beta } =  \ell_{\left(a\right)}^{\alpha }\ell _{\left(b\right)}^{\beta}\eta^{ab} = \left(\begin{array}{ccc} {a_{11} } & {a_{12} {\rm exp}\left(-u^{3} \right)} & {a_{13} } \\ {a_{12} {\rm exp}\left(-u^{3} \right)} & {a_{22} {\rm exp}\left(-2u^{3} \right)} & {a_{23} {\rm exp}\left(-u^{3} \right)} \\ {a_{13} } & {a_{23} {\rm exp}\left(-u^{3} \right)} & {a_{33} } \end{array}\right)
\end{equation}

\noindent
Second version.
The operators ${\rm X} _{a} $ have the following  form:
$${\rm X} _{1} =p_{2} ,\quad {\rm X} _{2} =p_{3} ,\quad {\rm X} _{3} ={\rm A}_{3} p_{1} +{\rm B}_3p_{2} +{\rm R}_3 p_{3}.$$
From the equations of the system \eqref{1c}, after performing an admissible transformation, it follows that \quad
$${\rm A}_{3} =1,\quad {\rm B}_{3} =\ell _{0} \left(u^{0} \right), \quad {\rm R}_3=1.
$$ Therefore, the~set of group operators in this case is of the following  form:
\begin{equation} \label{6c}
\tilde{{\rm X}}_{3} = {\rm X} _{3}+\Delta {\rm X} _{3}, \quad {\rm X} _{1} ={\rm p} _{2}, \quad{\rm X} _{2} ={\rm p}_{3}, \quad {\rm X}_{3} ={\rm p} _{1} +u^{3}{\rm p}_{3}, \quad \Delta {\rm X} _{3} =\ell _{0} {\rm p}_{2}.
\end{equation}
Using the operators  ${\rm X} _{a}$,  one can find the vector fields $\ell _{\left(a\right)}^{\alpha} $ and the matrix $g^{\alpha \beta } =\ell _{\left(a\right)}^{\alpha } \ell _{\left(b\right)}^{\beta } \eta ^{ab} $.
To achieve this, the following matrices must be used:
$$\xi _{\left(a\right)}^{\alpha } =\left(\begin{array}{ccc} {0} & {1} & {0} \\ {0} & {0} & {1} \\ {1} & {0} & {u^{3} } \end{array}\right), \quad \xi _{\alpha }^{\left(a\right)} =\left(\begin{array}{ccc} {0} & {-u^{3} } & {1} \\ {1} & {0} & {0} \\ {0} & {1} & {0} \end{array}\right), \quad \xi _{\left(a\right),\gamma }^{\alpha } =\left(\begin{array}{ccc} {0} & {0} & {0} \\ {0} & {0} & {0} \\ {0} & {0} & {\delta _{\gamma }^{3} } \end{array}\right).
$$
The solution of  system \eqref{14},
 $$\ell^\alpha_{(a),\beta} = \delta^\alpha_3 \delta^1_\beta \ell^3_{(a)},  $$
 can be written as follows:
\begin{equation} \label{7c}
\ell _{\left(a\right)}^{\alpha } =\delta _{1}^{\alpha } \delta _{a}^{1} +\delta _{2}^{\alpha } \delta _{a}^{2} +{\rm exp}u^{1} \delta _{3}^{\alpha } \delta _{a}^{2}
\end{equation}
Therefore, the~matrix $g^{\alpha \beta } $ has the following form:
\begin{equation} \label{8c}
g^{\alpha \beta } =\begin{pmatrix}a_{11}  & a_{12}  & a_{13} \exp{u^1} \\ a_{12}  & a_{22} & {a_{23} \exp{u^1}} \\ a_{13}{\exp{u^1}} & a_{23} \exp{u^1} & a_{33} \exp{2u^1}\end{pmatrix}
\end{equation}
Let us represent the metric tensor  $\tilde{g}^{\alpha \beta }$ as
$$ \tilde{g}^{\alpha \beta }  = g^{\alpha \beta }+G^{\alpha \beta } $$
From the the Killing equations, it follows that
$$G^{\alpha \beta } = \dot{\ell }_{0}u^{1}(\delta _{1}^{\alpha } \delta _{2}^{\beta } +\delta _{2}^{\alpha } \delta _{1}^{\beta })$$
Then,
\begin{equation} \label{9c}
\tilde{g}^{ij} =\begin{pmatrix}a_{11}  & a_{12}  & \dot{\ell }_{0}u^{1} + a_{13} \exp{u^3} \\ a_{12}  & a_{22} & {a_{23} \exp{u^3}} \\\dot{\ell }_{0}u^{1} + {\exp{u^3} a_{13}} & a_{23} \exp{u^3} & a_{33} \exp{2u^3}\end{pmatrix}
\end{equation}

\quad

\noindent
Third version. The operators ${\rm X} _{a} $ are of the form  ${\rm X} _{1} ={\rm p} _{2}, {\rm X} _{2} =a_{2} {\rm p}_{1} +b_{2} {\rm p}_{2}$ 
 \linebreak  ${\rm X} _{3} = a_{3} {\rm p}_{1} +b_{3}{\rm p}_{2} +{\rm p}_{3} $. After~admissible transformations, the~functions $a_{3} , b_{3} $ are converted to zero. From~the equation $\left[{\rm X} _{2} {\rm X} _{3} \right]={\rm X} _{2} $, it follows that
$${\rm X} _{2} ={\rm p}_{1}\exp{-u^{3}}$$
Then, the set of group operators has the following  form:
\begin{equation} \label{10c}
{\rm X}_{1}= {\rm p}_{2}, \quad {\rm X} _{2} ={\rm p}_{1}\exp{-u^{3}}, \quad {\rm X} _{3} ={\rm p}_{3}.
\end{equation}
Using the matrices
 $$\xi _{\left(a\right)}^{\alpha } =\left(\begin{array}{ccc} {{\rm exp}\left(-u^{3} \right)} & {0} & {0} \\ {0} & {1} & {0} \\ {0} & {0} & {1} \end{array}\right), \quad \xi _{\alpha }^{\left(a\right)} =\left(\begin{array}{ccc} {{\rm exp}u^{3} } & {0} & {0} \\ {0} & {1} & {0} \\ {0} & {0} & {1} \end{array}\right),$$
 $$\quad \xi _{\left(a\right),\gamma }^{\alpha } =-\delta _{\gamma }^{3} {\rm exp}\left(-u^{3} \right)\left(\begin{array}{ccc} {1} & {0} & {0} \\ {0} & {0} & {0} \\ {0} & {0} & {0} \end{array}\right),$$
let us construct the system of Equation \eqref{14}:
\begin{equation} \label{11c}
\ell _{\left(a\right),1}^{1} =-\ell _{a}^{3} ,  \quad \ell _{\left(a\right),\alpha}^{p} =0
\end{equation}
and find the vector fields $\ell _{\left(a\right)}^{\alpha } $:
\begin{equation} \label{12c}
\ell _{\left(a\right)}^{\alpha } =\delta _{1}^{\alpha } (\delta _{a}^{1} -u^{1} \delta _{a}^{3})+\delta _{2}^{\alpha } (\delta _{a}^{2} +\delta _{3}^{\alpha } \delta _{a}^{3}).
\end{equation}
The matrix of tensor  ${g}^{\alpha \beta }= \ell _{\left(a\right)}^{\alpha }\ell _{\left(b\right)}^{\beta}\eta^{ab}$  has the following  form:
\begin{equation} \label{13c}{g}^{\alpha \beta }  =\left(\begin{array}{ccc} {a_{11} -2u_{1} a_{13} +u_{1}^{2} } & {a_{12} -u^{1} a_{23} } & {a_{13} -a_{23} u^{1} } \\ {a_{12} -u^{1} a_{23} } & {a_{22}} & {a_{23}} \\ {a_{13} -u^{1} a_{33} } & {a_{23} } & {a_{33} } \end{array}\right)
\end{equation}

\subsection{${Group \quad G_{3}(IV)}$}

The structural equations have the following  form:
\begin{equation} \label{1cc}\left[{\rm X} _{1} {\rm X} _{2} \right]=0, \quad \left[{\rm X} _{1} {\rm X} _{3} \right]={\rm X} _{1}, \quad   \left[{\rm X} _{2} {\rm X} _{3} \right]={\rm X} _{1} +{\rm X} _{2}.
\end{equation}

\noindent
First version. The group operators have the following form:
\begin{equation} \label{2cc}
{\rm X} _{1} =a_{1} {\rm p}_{1}, \quad {\rm X} _{2} = {\rm p}_{2},\quad {\rm X} _{3} =a_{3} {\rm p}_{1} + b_{3}{\rm p}_{2} +r_{3}{\rm p} _{3}.
\end{equation}
From the second equation of the system \eqref{1cc}, it follows that
 ${\rm X} _{1} ={\rm p} _{1} {\rm exp}\left(-u_{3} \right)$.
Let us substitute this in the last equation of the system \eqref{1cc} and perform an admissible transformation of coordinates. As~a result, we obtain the following:
\begin{equation} \label{3cc}
{\rm X} _{1} ={\rm p} _{1} {\rm exp}\left(-u^{3} \right), \quad {\rm X} _{2} ={\rm p} _{2}, \quad{\rm X} _{3} ={\rm p} _{3} +u^{2} \left({\rm p} _{1} {\rm exp}\left(-u^{3} \right)+{\rm p} _{2} \right)
\end{equation}
The matrices $\xi ^{\alpha }_{\left(a\right)},\xi _{\alpha }^{\left(a\right)}, \xi _{\alpha ,\gamma }^{\beta } $ have the following form:
 $$\xi _{\left(a\right)}^{\alpha } =\left(\begin{array}{ccc} {{\rm exp}\left(-u^{3} \right)} & {0} & {0} \\ {0} & {1} & {0} \\ {{\it u}_{2} {\rm exp}\left(-u^{3} \right)} & {u^{2} } & {1} \end{array}\right), \quad
\xi _{\alpha }^{\left(a\right)} =\left(\begin{array}{ccc} {{\it exp}u^{3} } & {0} & {0} \\ {0} & {1} & {0} \\ {-u^{2} } & {-u^{2} } & {1} \end{array}\right),
$$
$$
\xi _{\left(a\right),\gamma }^{\alpha } =\delta _{\gamma }^{2} \left(\begin{array}{ccc} {0} & {0} & {0} \\ {0} & {0} & {0} \\ {\rm exp}\left(-u^{3} \right) & {1} & {0} \end{array}\right)-\delta _{\gamma }^{3} {\rm exp}\left(-u^{3} \right)\left(\begin{array}{ccc} {1} & {0} & {0} \\ {0} & {0} & {0} \\ {u^{2} } & {0} & {0} \end{array}\right)$$
Using them, let us obtain the set of Equation \eqref{15}:
$$
\ell^1_{(a),1} = -\ell^3_{(a)}, \quad \ell^2_{(a),3} = \ell^2_{(a)}, \quad \ell^1_{(a),3} = \ell^2_{(a)}{\rm exp(-u^3)}, \quad \ell^3_{(a),\alpha} =0.
$$
The solution can be written in the following  form:
\begin{equation} \label{4cc}
\ell _{a}^{\alpha } =\delta _{1}^{\alpha } \left(\delta _{a}^{1} -u^{1} \delta _{a}^{3} +u^{3} \delta _{a}^{2} \right)+{\rm exp}u^{3}\delta _{2}^{\alpha } \delta _{a}^{2} +\delta _{3}^{\alpha } \delta _{a}^{3}. \quad
\end{equation}
 Let us find matrix  $g^{\alpha \beta }= \ell _{\left(a\right)}^{\alpha }\ell _{\left(b\right)}^{\beta}\eta^{ab}:$
\begin{equation} \label{5cc}
g^{\alpha \beta }=\left(\begin{array}{ccc} {a_{11}+2\left(u^{3} a_{12} -u^{1} a_{13} \right)+\left(a_{33} {u^{1}}^{2} +a_{22} {u^{3}}^{2} -2u^{1} a_{23} \right)} & {g^{12}} & {g^{13}} \\ {\left(a_{12} -u^{1} a_{23} +u^{3} a_{22} \right){\rm exp}u^{3}} & a_{22} \exp{2u^{3}} & {g^{23} } \\ {a_{13} -u^{1} a_{33} +u^{3} a_{23} } & {a_{23}\exp u^{3} } & {a_{33} } \end{array}\right)
\end{equation}

\quad

\noindent
Second version. The operators ${\rm X} _{a} $ have the following  form: \quad $${\rm X} _{1} =p_{2} ,\quad {\rm X} _{2} =p_{3} ,\quad {\rm X} _{3} ={\rm A}_{3} p_{1} +{\rm B}_3p_{2} +{\rm R}_3 p_{3}. $$
From structural Equation \eqref{1cc}, it follows that
\begin{equation} \label{6cc}
{\rm X} _{1} =p_{2} ,\quad {\rm X} _{2} =p_{3},\quad {\rm X}_{3} =p_{1} +(u^{2} + u^{3} )p_{2}+u^{3}p_{3}.
\end{equation}
To obtain Equation \eqref{15}, we use the following matrices:
 $$\xi _{\left(a\right)}^{\alpha } =\begin{pmatrix} {1} & {(u^{2} + u^{3} )} & {u^3} \\ {0} & {1} & {0} \\ {0} & {0} & {1} \end{pmatrix}, \quad
\xi _{\alpha }^{\left(a\right)} =\begin{pmatrix} {1} & {-(u^{2} + u^{3} )} & {-u^3} \\ {0} & {1} & {0} \\ {0} & {0} & {1} \end{pmatrix}, \quad
\xi _{\left(a\right),\gamma }^{\alpha } =\begin{pmatrix}  {0} & {(\delta _{\gamma }^{2}+\delta _{\gamma }^{3} )} & {\delta _{\gamma }^{3} } \\ {0} & {0} & {0} \\ {0} & {0} & {0} \end{pmatrix}, $$
Equations \eqref{15} have the following form:
\begin{equation} \label{7cc}
\ell _{\left(a\right),\beta }^{\alpha } =\left(\begin{array}{ccc} {0} & {\ell _{\left(a\right)}^{2}+\ell _{\left(a\right)}^{3}} & {\ell _{\left(a\right)}^{3} } \\ {0} & {0} & {0} \\ {0} & {0} & {0} \end{array}\right)\quad
\Rightarrow \quad
\ell _{\left(a\right)}^{\alpha } =\delta _{1}^{\alpha } \delta _{a}^{1} +(\delta _{2}^{\alpha } (\delta _{a}^{2} + u^1\delta _{a}^{3}) + \delta _{3}^{\alpha } \delta_{a}^{3})\exp u^{1}
\end{equation}
Let us find matrix  $g^{\alpha \beta }= \ell _{\left(a\right)}^{\alpha }\ell _{\left(b\right)}^{\beta}\eta^{ab} $:
\begin{equation} \label{8cc}
g^{\alpha \beta } =\begin{pmatrix} a_{11} &(a_{12} +a_{13}u^1 )\exp u^1 & a_{13} \exp u^{1} \\ (a_{12} +a_{13}u^1)\exp u^1 & (a_{22} +2a_{23}u^1 + a_{33}{u^1}^2)\exp 2u^1 & (a_{23}+a_{33})\exp2{u}^{3} \\ a_{13}\exp u^{1} &(a_{23}+a_{33})\exp2{u}^{1}  & a_{33} \exp2u^{1}\end{pmatrix}
\end{equation}

\noindent
Third version. The operators ${\rm X} _{a} $ have the following form:
 $${\rm X} _{1} ={\rm p} _{2}, \quad {\rm X} _{2} =a_{2} {\rm p} _{1} +b_{2} {\rm p} _{2}, \quad {\rm X} _{3} ={\rm A} _{3} {\rm p} _{1} +{\rm B} _{3,} {\rm p} _{2} +R_{3} {\rm p} _{3} $$
Let us substitute this in equation $\left[{\rm X} _{1} {\rm X} _{3} \right]={\rm X} _{1} +{\rm X} _{2} $. As a result, we obtain the following:  $ {\rm B}_3 =u^2,  {\rm R}_3=1.$  From the last equation of system \eqref{1cc}, it follows that
\begin{equation}\label{9cc}
{\rm X}_{1} = {\rm p}_{2}, \quad {\rm X}_{2} = {\rm p}_{1}\exp{(-u^3)} - {\rm p}_{2} u^3, \quad {\rm X}_{3} = {\rm p}_{3} +  {\rm p}_{2} u^2.
\end{equation}
Using the following matrices,
\vspace{-6pt}
$$\xi _{\left(a\right)}^{\alpha } =\begin{pmatrix}\exp(-u^{3}) & {-u_{3} } & {0} \\ {0} & {1} & {0} \\ {0} & {u^{2} } & {1} \end{pmatrix},
 \quad \xi _{\alpha }^{\left(a\right)} =\begin{pmatrix} \exp u^{3} & {u^{3} {\rm exp}u^{3} } & {0} \\ {0} & {1} & {0} \\ {0} & {-u^{2} } & {1} \end{pmatrix},$$
 $$\xi _{\left(a\right),\gamma }^{\alpha } =\begin{pmatrix}-\delta_{\gamma }^{3} \exp(-u_3) & {-\delta_{\gamma }^{3}} & {0} \\ {0} & {0} & {0} \\ {0} & {\delta _{\gamma }^{2} } & {0} \end{pmatrix}.$$
one can obtain the system of Equation \eqref{14}:
 $$\ell _{\left(a\right),1}^{1} =-\ell _{\left(a\right)}^{3}, \quad \ell _{\left(a\right),1}^{2} =-\ell _{\left(a\right)}^{3} {\rm exp}\left(u^{3} \right), \quad \ell _{\left(a\right),3}^{2} =\ell _{\left(a\right)}^{2}, \quad \ell _{\left(a\right),2}^{\alpha} =0, \quad \ell _{\left(a\right),3}^{1} = 0. $$
 The solution has the following form:
\begin{equation} \label{10cc}
\ell _{\left(a\right)}^{\alpha } =\delta _{1}^{\alpha } \left(\delta _{a}^{1} +u^{1} \delta _{a}^{3} \right)+\delta _{2}^{\alpha } \left(\delta _{a}^{2}+\delta _{a}^{3} u^{1} \right)\exp u^3 -\delta _{3}^{\alpha } \delta _{a}^{3}, \quad
\end{equation}
Let us find the matrix of the metric tensor $g^{\alpha \beta } $:
\begin{equation} \label{11cc}
g^{\alpha \beta } =\left(\begin{array}{ccc} {a_{11} -2u_{1} a_{13} +{u^{1}}^{2} a_{33} } & {a_{12} -u^{1} \left(a_{33} +a_{23} \right)+{u^{1}}^{2} a_{33} } & {a_{13} -u^{1} a_{33} } \\ {a_{12} -u^{1} \left(a_{33} +a_{23} \right)+{u^{1}}^{2} a_{33} } & {a_{22}^{2} -2u^{1} a_{23} +{u^{1}}^{2} a_{33} } & {a_{23} -u^{1} a_{33} } \\ {a_{13} -u^{1} a_{33} } & {a_{23} -u^{1} a_{33} } & {a_{33} } \end{array}\right)
\end{equation}

\quad

\subsection{${Group \quad G_{3}(V)}$}

The structural equations have the form
\begin{equation} \label{1aa}\left[{\rm X} _{1} {\rm X} _{2} \right]=0, \quad \left[{\rm X} _{1} {\rm X} _{3} \right]={\rm X} _{1}, \quad \left[{\rm X} _{2} {\rm X} _{3} \right]={\rm X} _{2}. \end{equation}

\noindent
First version. From structural equation 
\begin{equation*}
\begin{array}{ccc} {{\rm X} _{1} =p_{1} {\rm exp}u_{3} ,} & {{\rm X} _{2} =p_{2} ,} & {{\rm X} _{3} =-p_{3} +u^{2} p_{2} } \end{array}
\end{equation*}
This version is equivalent to Petrov's result (\cite{16}, (f. 25.24), p. 163) (after variable transformation: $\widetilde{u}^{1} =u^{1} {\rm exp}u_{3} $). On~the other hand, it is a partial case of the third variant, which is considered below. Therefore, it should be excluded from the classification.

\noindent
Second version.
 ${\rm X} _{1} ={\rm p} _{2},  {\rm X} _{2} ={\rm p} _{3} $. From the structural equations, it follows (after using the admissible transformations of variables) that
 $${\rm X} _{3} ={\rm p} _{1} +u^{2} {\rm p} _{2} +u^{3} {\rm p} _{3}. $$
Using the following matrices,
$$\xi _{\left(a\right)}^{\alpha } =\begin{pmatrix}0 & 1 & {0} \\ {0} & {0} & {1} \\ {1} & {u^{2} } & {u^{3} } \end{pmatrix},
 \quad \xi _{\alpha }^{\left(a\right)} =\begin{pmatrix} {-u^{2} } & {-u^{3}}  & {1} \\ {1} & {0} & {0} \\ {0} & {1} & {0} \end{pmatrix},\quad
\xi _{\left(a\right),\gamma }^{\alpha } =\begin{pmatrix}0 & {0} & {0} \\ {0} & {0} & {0} \\ {0} & {\delta _{\gamma }^{2} } & \delta _{\gamma }^{2} \end{pmatrix},$$
one can obtain the system of Equation \eqref{14}:
 $$\ell _{\left(a\right),\alpha }^{1} =0, \quad \ell _{\left(a\right),  \quad \alpha }^{2} =\delta _{\alpha }^{1} \ell _{\left(a\right)}^{2},  \quad \ell _{\left(a\right),\alpha }^{3} =\delta _{\alpha }^{1} \ell _{\left(a\right)}^{3} $$
The solutions have the following form:
$$\ell _{\left(a\right)}^{\alpha } =\delta _{a}^{1} \delta _{1}^{\alpha } +\exp u^{1} (\delta _{2}^{a} \delta _{a}^{2} +\delta _{3}^{\alpha } \delta _{a}^{3})$$
Let us find components  of the matrix $g^{\alpha \beta } = \ell _{\left(a\right)}^{\alpha }\ell _{\left(b\right)}^{\beta}$ :
\begin{equation} \label{2aa}
g^{\alpha \beta } =\left(\begin{array}{ccc} {a_{11} } & {a_{12}\exp u^1} & {a_{13}\exp u^1} \\ {a_{12}\exp u^1} & {a_{22}}\exp 2u^1 & {a_{23}\exp 2u^1} \\ {a_{13}\exp u^1} & {a_{23}\exp 2u^1} & {a_{33}\exp 2u^1} \end{array}\right),
\end{equation}

\noindent
Third version  ${\rm X} _{1} ={\rm p} _{2},  {\rm X}_{2} =a_{2} {\rm p}_1 +b_2 {\rm p}_2$.  From structural Equation \eqref{1aa}, it follows that
$${\rm X} _{2} ={\rm p } _{1} {\rm exp}u^{3} +\ell_{0} {\rm p }_{2}, \quad {\rm X} _{3} ={\rm p }_{3} + u^2{\rm p }_{2}.
$$

\noindent
 Using the following matrices,
 $$\xi _{\left(a\right)}^{\alpha } =\begin{pmatrix}\exp u^{3} & 0 & {0} \\ {0} & {1} & {0} \\ {0} & {u^2} & {1} \end{pmatrix},
 \quad \xi _{\alpha }^{\left(a\right)} =\begin{pmatrix} \exp (-u^{3}) & {0} & {0} \\ {0} & {1} & {0} \\ {0} & {u^{2} } & {1} \end{pmatrix},$$
 $$\xi _{\left(a\right),\gamma }^{\alpha } =\begin{pmatrix}\delta_{\gamma }^{3} & {0} & {0} \\ {0} & {0} & {0} \\ {0} & {\delta _{\gamma }^{2} } & {0} \end{pmatrix},$$
one can obtain the system of Equation \eqref{14}:
 $$\ell _{\left(a\right),1}^{1} =\ell _{\left(a\right)}^{3}, \quad \quad \ell _{\left(a\right),3}^{2} =\ell _{\left(a\right)}^{2}, \quad \ell _{\left(a\right),2}^{\alpha} = 0,$$
 $$\quad \ell _{\left(a\right),\alpha}^{3} = 0,\quad \ell _{\left(a\right),3}^{\alpha} = 0,
 \quad \ell _{\left(a\right),p}^{1} = 0, \quad \quad \ell _{a,1}^{2}=0.$$
 The solution can be represented in the following form:
\begin{equation} \label{4aa}
\ell _{\left(a\right)}^{\alpha } =\delta _{1}^{\alpha } \left(\delta _{a}^{1} +u^{1} \delta _{a}^{3} \right)+\delta _{2}^{\alpha } \delta _{a}^{2}\exp u^3 +\delta _{3}^{\alpha } \delta _{a}^{3}
\end{equation}
Then, the components  of matrix $g^{\alpha \beta } =\ell _{\left(a\right)}^{\alpha }\ell _{\left(b\right)}^{\beta }\eta^{\alpha\beta}  $ will be as follows:
\begin{equation} \label{5aa}
g^{\alpha \beta } =\left(\begin{array}{ccc} {a_{11} + 2u^1a_{13} +a_{33}{u^1}^2} & (a_{12}+ u^1 a_{13}) \exp u^3 & {a_{13}+ u^1 a_{33}} \\ (a_{12}+ u^1 a_{13}) \exp u^3  & {a_{22}\exp 2u^3} & {a_{23}\exp u^3} \\ {a_{13}+ u^1 a_{33}} & {a_{23}\exp u^3} & {a_{33}} \end{array}\right).
\end{equation}
The tensor $\tilde{g}^{\alpha \beta } $ can be represented as follows (Killing equations have been used):
\[\tilde{g}^{\alpha \beta} = g^{\alpha \beta } +\ell_{0,0} u^{1} \left(\delta _{1}^{\alpha } \delta _{2}^{\beta } +\delta _{1}^{\beta } \delta _{2}^{\alpha } \right)\exp{u^3}\]

\subsection{${Group \quad G_{3}(VI)}$}

The structural equations have the following  form:
\begin{equation}\label{5.1}
\left[ {\rm X} _{1} {\rm X} _{2}\right]=0,\quad \left[{\rm X} _{2} {\rm X} _{3}\right] \quad =q {\rm X} _{2} ,\quad \left[{\rm X} _{1} {\rm X} _{3}\right] ={\rm X} _{1}.
 \end{equation}

\noindent
First version.
${\rm X} _{1} =p_{1} {\rm exp}\left(-u^{3} \right),  {\rm X} _{2} =p_{2} $. From the structural equation, it follows that
\begin{equation} \label{5.7}
{\rm X} _{1} =p_{1} {\rm exp}\left(-u^{3} \right), \quad {\rm X} _{2} ={\rm p}_{2}, \quad {\rm X}_{3} ={\rm p}_{3} +qu^{2} {\rm p}_2.
\end{equation}
The set corresponds to the Petrov set (\cite{16}, (f. 25.24), p. 163, when substituting  $\widetilde{u}^{1} =u^{1} \exp u^{3} $). Let us find the vectors $\ell _{\left(a\right)}^{\alpha } $ using the following matrices:
 $$\xi _{\left(a\right)}^{\alpha } =\begin{pmatrix}{\exp(-u^{3} )} & {0} & {0} \\ {0} & {1} & {0} \\{0} & {qu^{2} } & {1} \end{pmatrix}, \quad \xi _{\alpha }^{\left(a\right)} =\begin{pmatrix}{\exp u^{3} } & {0} & {0} \\ {0} & {1} & {0} \\{0} & {-qu^{2} } & {1} \end{pmatrix}, \quad
\xi _{\left(a\right),\gamma }^{\alpha } =\begin{pmatrix}-\exp(-u^{3}) \delta_{\gamma }^{3}  & {0} & {0} \\ {0} & {0} & {0} \\{0} & q\delta _{\gamma }^{2}  & {0} \end{pmatrix}.$$
The equations of \eqref{14} take the form
 $$\ell _{\left(a\right),1}^{1} =-\ell _{\left(a\right)}^{3}, \quad \ell _{\left(a\right),3}^{2} =q\ell^{2}_{(a)}, \quad  \ell _{\left(a\right),2}^{\alpha } =0, \quad \ell _{\left(a\right),\alpha }^{3} =0, \quad \ell _{\left(a\right),1}^{2} =0.$$
The solution can be written through the operators  $Y_{a} =\ell _{\left(a\right)}^{\alpha } p_{\alpha } $  in the following form:
\begin{equation} \label{5.7}
Y_{1} =p_{1}, \quad Y_{2} =p_{2} {\rm exp}\left(qu^{3} \right), \quad Y_{3} =p_{3} -u^{1} p_{1} \Rightarrow \ell _{\left(a\right)}^{\alpha } =\delta_{1}^{\alpha } \delta_{a}^{1} +\delta _{2}^{\alpha } \delta _{a}^{2} +{\rm exp}u^{3} \delta _{3}^{\alpha } \delta _{a}^{2}
\end{equation}
$$\Rightarrow \ell _{\left(a\right)}^{\alpha } =\delta _{1}^{\alpha } \delta _{a}^{1} +\delta _{2}^{\alpha } \delta _{a}^{2} +{\rm exp}u^{3} \delta _{3}^{\alpha } \delta _{a}^{2}.$$
Let us provide the elements of the matrix $g^{\alpha \beta } $:
 $$ g^{11} =\begin{pmatrix}a_{11}+2u^{1} a_{13} +{u^{1}}^{2}a_{33} & (a_{12}+u^{1}a_{23} )\exp qu^{3}& a_{13} +u^{1}a_{33}\\(a_{12}+u^{1}a_{23} )\exp qu^{3} & a_{22} \exp 2qu^{3}& a_{23}\exp qu^3\\a_{13} + u^{1} a_{33}& a_{23}\exp qu^3 & a_{33} \end{pmatrix}.
 $$

\noindent
Second version.  ${\rm X} _{1} =p_{2} ,  {\rm X} _{2} =p_{3}.$
 From the structural equations we immediately obtain
$${\rm X} _{3} =p_{1} +u^{2} p_{2} +qu^{3} p_{3}. $$
Let us find the vector fields $\ell _{\left(a\right)}^{\alpha } $. Using the following matrices,
\[\xi _{\left(a\right)}^{\alpha } =\left(\begin{array}{ccc} {1} & {u^{2} } & {qu^{3} } \\ {0} & {1} & {0} \\ {0} & {0} & {1} \end{array}\right), \quad \xi _{\alpha }^{\left(a\right)} =\left(\begin{array}{ccc} {1} & {-u_{2} } & {-qu^{3} } \\ {0} & {1} & {0} \\ {0} & {0} & {1} \end{array}\right), \quad\xi _{\left(a\right),\gamma }^{\alpha } =\left(\begin{array}{ccc} {0} & {\delta _{\gamma }^{2} } & {q\delta _{\gamma }^{3} } \\ {0} & {0} & {0} \\ {0} & {0} & {0} \end{array}\right),\]
one can obtain Equation \eqref{14} in the following  form:
\begin{equation} \label{5.1}
\ell _{\left(a,\alpha\right)}^{1} =\ell _{\left(a,2\right)}^{\alpha } =\ell _{\left(a,3\right)}^{\alpha } =0, \quad \ell _{\left(q,1\right)}^{2} =\ell_{\left(q\right)}^{2}, \quad \ell_{\left(a,1\right)}^{3} =q\ell_{\left(a\right)}^{3}
\end{equation}
Let us find solution of system \eqref{5.1}:
\begin{equation} \label{5.2}
Y_{1} =p_{1}, \quad Y_{2} =p_{2} {\exp}u^{1}, \quad Y_{3} =p_{3}{\exp}(qu^{1}) \Rightarrow
\end{equation}
$$\ell _{\left(a\right)}^{1} =\delta _{a}^{1},\quad \ell _{\left(a\right)}^{2} =\delta _{a}^{2} {\rm exp}u^{1}, \quad \ell _{\left(a\right)}^{3} =\delta _{a}^{3}{\rm expq}u^{1}$$
Matrix $g^{\alpha \beta }=\ell _{\left(a\right)}^{\alpha }\ell _{\left(b\right)}^{\beta }\eta^{\alpha\beta} $ has the following  form:
\begin{equation} \label{5.3}
g^{\alpha \beta } =\left(\begin{array}{ccc} {a_{11} } & {a_{12} {\rm exp}u^{1} } & {a_{13} {\rm exp}{\it q}u^{1} } \\ {a_{12} {\it exp}u^{1} } & {a_{22} {\it exp2}u^{1} } & {a_{23} {\it exp}\left(q+1\right)u^{1} } \\ {a_{13} {\it expq}u^{1} } & {a_{23} {\it exp}\left(q+1\right)u^{1} } & {a_{33} {\it exp2q}u^{1} } \end{array}\right)
\end{equation}

\noindent
Third version.
  The operators ${\rm X} _{a} $ have the following  form:
 $${\rm X} _{1} ={\rm p}_{2} , \quad {\rm X} _{2} =a_{2} {\rm p}_{1} +b_{2} {\rm p}_{2} , \quad {\rm X} _{3} ={\rm A} _{3} {\rm p}_{1} +{\rm B} _{3}{\rm p}_{2} +R_{3}{\rm p}_{3}. $$
From the structural equations, it follows that
 $${\rm A} _{3,2}{\rm p}_{1} +{\rm B} _{3,2}{\rm p}_{2} +R_{3,2} {\rm p}_{3} ={\rm p}_{2}.$$
Hence, $${\rm A} _{3} =a_{3},\quad {\rm B} _{3} =u^{2} +b_{3},\quad R_{3} =r_{3}.\quad $$ Since ${\rm r}_{3} \ne 0$, the~function ${\rm r}_{3}$ can be be turned  to $(-1)$. Then, the functions  $a_{3}, b_{3} $  are reversed to zero by admissible coordinate transformations. The~operator ${\rm X}_{3} $ will take the form ${\rm X} _{3} =-p_{3} +u^{2} p_{2} $.
Let us substitute ${\rm X} _{2} $ and ${\rm X} _{3} $ into the last structural equation. The~result is as follows:
 $$ b_{2} p_{2} +a_{2,3} p_{1} +b_{2,3} =q(a_{2} p_{1} +b_{2} p_{2})$$
Hence,  $a_{2} =a_{0}\exp(qu^{3}), b_{2} =b_{0} \exp((q+1)u^{3}). $
By using the admissible  coordinate transformation, one can turn $a_{0} ,b_{0} $ to unity
. Finally, the~set has the following form:
\begin{equation} \label{5.4}
{\rm X} _{1} ={\rm p}_{2}, \quad {\rm X} _{2} ={\rm}p_{1} \exp(qu^{3}) +{\rm p}_{2} \exp((q+1)u^{3}), \quad {\rm X} _{3} =-{\rm p}_{3} +u^{2} {\rm p}_2.
\end{equation}
The set differs from Petrov's set (\cite{16}, (f. 25.32), p. 165) by the transformation of the coordinate $u^{1} $.

Using the following  matrices,
$$\xi _{\left(a\right)}^{\alpha } =\left(\begin{array}{ccc} {{\rm exp}\left(u^{3} q\right)} & {{\rm exp}\left(q+1\right)u^{3} } & {0} \\ {0} & {1} & {0} \\ {0} & {u^{2} } & {-1} \end{array}\right),
 \quad \xi _{\alpha }^{\left(a\right)} =\left(\begin{array}{ccc} {{\rm exp}\left(-qu^{3} \right)} & {-{\rm exp}u^{3} } & {0} \\ {0} & {1} & {0} \\ {0} & {u_{2} } & {-1} \end{array}\right),
$$
$$\quad \xi _{\left(a\right),\gamma }^{\alpha } =\left(\begin{array}{ccc} {\delta _{\gamma }^{3} {\it q}{\it exp}\left(u^{3} q\right)} & {\delta _{\gamma }^{3} \left(q+1\right){\it exp}\left(u^{3} \left(q+1\right)\right)} & {0} \\ {0} & {0} & {0} \\ {0} & {\delta _{\gamma }^{2} } & {0} \end{array}\right)$$
we obtain the system of Equation \eqref{14} in the following  form: $$\ell _{\left(a\right),1}^{1} =q\ell _{\left(a\right)}^{3},\quad \ell _{\left(a\right),1}^{2} =\left(q+1\right)\ell _{\left(a\right)}^{3} ,\quad \ell _{\left(a\right)}^{2},$$
 $$\ell _{(a)}^{3} =c_{a}^{3}=const ,\quad \ell _{\left(a\right),2}^{\alpha } =\ell _{\left(a\right),3}^{1} =0.$$
The solution can be written in the following form:
\begin{equation} \label{5.5}
{\rm Y}_{1} =p_{1}, \quad {\rm Y}_{2} ={\rm p}_{2} {\exp}u^{3}, \quad {\rm Y}_{3} =p_{3}+u^1({\rm p}_{2}(1-q){\exp}u^{3}-q{\rm p}_{1}) \Rightarrow
\end{equation}
$$
\ell _{\left(a\right)}^{\alpha } =\delta _{1}^{\alpha }( \delta _{a}^{1} - qu^{1} \delta _{a}^{3}) +{\exp}u^{3}\delta _{2}^{\alpha}( \delta _{a}^{2}+u^1(1-q)\delta _{a}^{3}) + \delta _{3}^{\alpha } \delta _{a}^{3}
$$
Using vectors $\ell _{\left(a\right)}^{\alpha } $, let us find
elements of
~the matrix
$g^{\alpha \beta } =\ell _{\left(a\right)}^{\alpha }\ell _{\left(b\right)}^{\beta }\eta^{\alpha\beta} =  $
\begin{equation} \label{5.6}
\begin{pmatrix} a_{11} -2qu^{1} a_{13}+q^2{u^{1}}^{2} a_{33}&\exp u_{3}(a_{12} -u^{1}(a_{13} (q-1)+qa_{23})-q(q-1){u^{1}}^{2} a_{33})     & g^{13}\\ g^{12} & \exp 2u^{3} (a_{22} +2(q-1)a_{23} u^{1} +(q-1)^{2}{u^1}^2 a_{33})& g^{23}\\
a_{13}-qu^{1} a_{33}&\exp u^{3}(a_{23} +(q-1)u^{1} a_{33}) &a_{33}.  \end{pmatrix}
\end{equation}

\subsection{${Group \quad G_{3}(VII)}$}

The structural 
 equations have the following  form:
\begin{equation}\label{6.1}
\begin{array}{ccc} {\left[{\rm X} _{1} {\rm X} _{2} \right]=0} & {\left[{\rm X} _{1} {\rm X} _{3} \right]={\rm X} _{2} } & {\left[{\rm X} _{2} {\rm X} _{3} \right]=2\cos\alpha {\rm X} _{2} -{\rm X} _{1} } \end{array}.
\end{equation}
Here, I denote
~$q=2\cos\alpha =const$.

\noindent
First version.
Obviously, the~first variant cannot be realized, since the first two equations of the structure imply  ${\rm X}_1={\rm p}_1$.   In~this case, ${\rm X}_1$ commutes with ${\rm X}_p$, which is~impossible.

\noindent
Second version.
 ${\rm X} _{1} =p_{2},  {\rm X} _{2} =p_{3},  {\rm X} _{3} ={\rm A} p_{1} +{\rm B} p_{2} +Rp_{3} $. From~structural Equation \eqref{6.1}, it follows that
\begin{equation}\label{7.1a}
 {\rm X} _{1} =p_{2}, \quad {\rm X} _{2} =p_{3}, \quad  {\rm X} _{3} =p_{1} +u^{3}. (2\cos{\alpha} p_{3} -p_{2})+u^{2} p_{3}
\end{equation}
Let us find the vector fields $\ell _{\left(a\right)}^{\alpha }. $ Using the following matrices,
\[\xi _{\left(a\right)}^{\alpha } =\left(\begin{array}{ccc} {1} & {-u^{3} } & {2u^{3} \cos{\alpha} +u^{2} } \\ {0} & {1} & {0} \\ {0} & {0} & {1} \end{array}\right),\xi _{\alpha }^{\left(a\right)} =\left(\begin{array}{ccc} {1} & {u^{3} } & {-\left(u^{2} +2 u^{3}\cos{\alpha} \right)} \\ {0} & {1} & {0} \\ {0} & {0} & {1} \end{array}\right),\]
\[\xi _{\left(a\right),\gamma }^{\alpha } =\left(\begin{array}{ccc} {0} & {-\delta _{\gamma }^{3} } & {\delta _{\gamma }^{2} +2\delta _{\gamma }^{3} \cos{\alpha}} \\ {0} & {0} & {0} \\ {0} & {0} & {0} \end{array}\right),\]
one can obtain the system of Equation \eqref{14} in the following  form:
\begin{equation} \label{7.2}\ell _{\left(a\right),1}^{1} =0, \quad \ell _{\left(a\right),1}^{2} =-\ell _{\left(a\right)}^{3}, \quad\ell _{\left(a\right),1}^{3} =\ell _{\left(a\right)}^{2} +2\ell _{a}^{3} \cos{\alpha}, \quad \ell _{\left(a\right),2}^{\alpha } =\ell _{\left(a\right),3}^{\alpha } =0.\end{equation}
Let us represent the solutions in the form  $Y_{a} =\ell _{a}^{\alpha } p_{\alpha } $:
\vspace{-6pt}
\begin{equation}\label{7.3}
Y_{1} =p_{1},\quad Y_{2} =\left(p_{2} \sin p-p_{3} \sin \left(p+\alpha \right)\right)\exp q,\quad Y_{3} =\left(p_{2} \cos{p}+p_{3} \cos\left(p+\alpha \right)\right)\exp q \Rightarrow
\end{equation}
\begin{equation} \label{6.4}
\ell _{\left(a\right)}^{\alpha } =\delta _{1}^{\alpha } \delta _{a}^{1} + \exp q (\delta _{2}^{\alpha }(\delta_{a}^{2}\sin p + \delta_{a}^{3}\cos p ) + \delta _{3}^{\alpha }(\delta_{a}^{2}\sin(p+\alpha) + \delta_{a}^{3}\cos(p+\alpha))),
\end{equation}
\noindent
where $q=u^{1} \cos{\alpha } ,  p=u^{1} \sin \alpha $.
Using $\ell _{\left(a\right)}^{\alpha } $ from \eqref{6.4}, one can find the elements of the matrix $g^{\alpha \beta }= \ell _{\left(a\right)}^{\alpha }\ell _{\left(b\right)}^{\beta }\eta^{ab}$:

\begin{equation} \label{7.4}
\left\{\begin{array}{c} {g^{11} =a_{11} ,g^{12} =\exp q\left(a_{12} \sin p+a_{13} \cos{p}\right),g^{13} =-\exp q\left(a_{12}\sin \left(p+\alpha \right) +a_{13} \cos\left(p+\alpha \right)\right)} \\ {g^{23} =-\exp 2q\left(\frac{\left(a_{22} +a_{33} \right)}{2} \cos\alpha +\frac{\left(a_{33}-a_{22} \right)}{2} \cos\left(2p+\alpha \right)+a_{23} \sin \left(2p+\alpha \right)\right)} \\ {g^{22} =\exp 2q\left(\frac{a_{22} +a_{33} }{2} +\frac{\left(a_{33} - a_{22} \right)}{2} \cos2p+a_{23} \sin 2p\right),} \\ {g^{33} =-\exp 2q\left(\frac{a_{22} +a_{33} }{2} +\frac{\left(a_{33}-a_{22} \right)}{2} \cos2\left(p+\alpha \right)+a_{23} \sin 2\left(p+\alpha \right)\right)} \end{array}\right.
\end{equation}

\noindent
Third version.
The operators ${\rm X} _{a} $ have the following  form:
 $${\rm X} _{1} =p_{2} ,\quad {\rm X} _{2} =a_{2} p_{1} +b_{2} p_{2} ,\qquad {\rm X} _{3} ={\rm A} p_{1} +{\rm B} p_{2} +Rp_{3} $$
From equation.
,~after performing admissible transformations $\left[{\rm X} _{1} {\rm X} _{3} \right]={\rm X} _{1} $, it follows that \linebreak ${\rm X} _{3} =p_{3} +u^{2} p_{2} $. Let us substitute it into the equation $\left[{\rm X} _{2} {\rm X} _{3} \right]=2\cos\alpha {\rm X} _{2} -{\rm X} _{1} $. As~a result, we obtain a condition on the functions $a_{2}, b_{2} $:
 $$a_{2,3} =a_{2} \left(b_{2} -2\cos\alpha \right),\quad b_{2,3} =\left(b_{2} -\cos\alpha \right)^{2} +\sin ^{2} \alpha. $$
From here, it follows that
$$a_{2} =\frac{\exp \left(-q\right)}{\cos p} ,\qquad b_{2} =\cos\alpha +\sin \alpha \frac{\sin p}{\cos p}.$$
Here, $p=u^{3} \sin \alpha, q=u^{3} \cos\alpha $. Thus, the~set of operators ${\rm X} _{a} $ is reduced to the following form:
\begin{equation}\label{7.9a}
 {\rm X} _{1} ={\rm p}_{2} ,\quad {\rm X} _{3} ={\rm p}_{3} +u^{2} {\rm X} _{2} ,\quad {\rm X} _{2} ={\rm p}_{1}\frac{\exp - q}{\cos p} +{\rm p}_{2} \frac{\cos \left(p-\alpha \right)}{\cos p} .
 \end{equation}
Let us find the vector fields $\ell _{\left(a\right)}^{\alpha } $ using the following matrices:
$$\xi _{\left(a\right)}^{\alpha } =\begin{pmatrix} {0} & {1} & {0} \\ {\exp \left(q\right)} & {\frac{\cos\left(p-\alpha \right)}{\cos p} } & {0} \\ {u_{2} \frac{\exp \left(q\right)}{\cos p} } & {u_{2} \frac{\cos\left(p-\alpha \right)}{\cos p} } & {1} \end{pmatrix},\quad
\xi _{\alpha }^{\left(a\right)} =\begin{pmatrix}{-\exp \left(-q\right)\cos\left(p-\alpha \right)} & {\cos p\exp \left(-q\right)} & {0} \\ {1} & {0} & {0} \\ {0} & {-u_{2} } & {1} \end{pmatrix},$$
\[\xi _{\left(a\right),\gamma }^{\alpha } =\delta _{\gamma }^{2} \left(\begin{array}{ccc} {0} & {0} & {0} \\ {0} & {0} & {0} \\ {\frac{\exp q}{corp} } & {\frac{\cos \left(p-\alpha \right)}{\cos p} } & {0} \end{array}\right)+\delta _{\gamma }^{3} \left(\begin{array}{ccc} {0} & {0} & {0} \\ {-\frac{\exp q \cos\left(p+\alpha \right)}{\cos^{2} p} } & {\frac{\sin ^{2} \alpha }{\cos^{2} p} } & {0} \\ {-u_{2} \frac{\exp q \cos\left(p+\alpha \right)}{\cos^{2} p} } & {\frac{u_{2} \sin ^{2} \alpha }{\cos^{2} p} } & {0} \end{array}\right)\]
The system of Equation \eqref{14} has the following form:
\begin{equation}\label{7.9}
\ell _{\left(a\right),1}^{1} =\frac{\cos\left(p+\alpha \right)}{\cos p} \ell _{\left(a\right)}^{3} ,\quad \ell _{\left(a\right),1}^{2} =\frac{\sin ^{2} \alpha \exp \left(-q\right)}{\cos p} \ell _{a}^{3} ,\quad \ell _{\left(a\right),\alpha }^{3} =0,\quad \ell _{\left(a\right),2}^{\alpha }, =0
\end{equation}
\[\ell _{\left(a\right),3}^{1} =\frac{\exp q}{\cos p} ,\quad \ell _{\left(a\right),3}^{2} =\ell _{\left(a\right)}^{2} \frac{\cos\left(p-\alpha \right)}{\cos p}. \]
The solution can be represented in the following form:
\begin{equation} \label{30}
Y_{1} =p_{2} ,\quad Y_{2} =p_{1}\sin \alpha\frac{\sin p}{\cos p} +p_{2}\sin^2(\alpha)\frac{\exp q}{\cos p}, \quad  Y_{3} =p_{3} +u^{1} (Y_{2}-p_{1}\cos \alpha ).
\end{equation}
\begin{equation} \label{7.4}
\ell _{\left(a\right)}^{\alpha } =\delta _{1}^{\alpha } (\delta _{a}^{1}+\exp q (\delta ^{2}_{a} + u^2 \delta _{a}^{3}))+
\delta _{2}^{\alpha }\cos (p-\alpha)(\delta_{a}^{2} + \delta_{a}^{2}u^2) + \delta _{3}^{\alpha }\delta_{a}^{3}.
\end{equation}
Here are the components of the tensor $g^{\alpha \beta } = \ell _{\left(a\right)}^{\alpha}\ell _{\left(b\right)}^{\beta}\eta^{ab} $:

\begin{equation}\label{7.10}
 \left\{\begin{array}{c} {g^{11} =a_{11} +\frac{\cos^{2} \left(p+\alpha \right)}{\cos^{2} p} \left(a_{22} +2u^{1} a_{23} +{u^{1}}^{2} a_{33} \right)-\frac{2\cos\left(p+\alpha \right)}{\cos p} \left(a_{12} +u^{1} a_{13} \right)} \\ {g^{22} =\frac{\sin ^{4} \alpha \exp \left(2q\right)}{\cos^{2} p} \left(a_{22} +2u^{1} a_{23} +{u^{1}}^{2} a_{33} \right),\qquad g^{33} =a_{33} } \\ {g^{12} =\frac{\sin ^{2} \alpha \exp q}{\cos p} \left(a_{12} +u^{1} a_{13} -\frac{\cos\left(p+\alpha \right)}{\cos p} \left(a_{22} +2u^{1} a_{23} +{u^{1}}^{2} a_{33} \right)\right)} \\ {g^{13} =a_{13} \frac{\cos\left(p+\alpha \right)}{\cos p} \left(a_{23} +u^{1} a_{33} \right),\qquad g^{23} =\frac{\sin ^{2} \alpha \exp q}{\cos p} \left(a_{23} +u^{1} a_{33} \right)} \end{array}\right.
\end{equation}

\section{ Unsolvable~Groups}\label{sec5}

For unsolvable groups $G_3(VIII), G_3(IX) $, there is only the fourth version:
$$
{\rm X}_1 = {\rm p}_2.
$$
For group $G_3(IX) $, it does not matter which of the operators ${\rm X}_a$ have to be diagonalized. In~the case of group $G_3(VIII)$, following Petrov, we choose the operator ${\rm X}_1$ as the diagonalized operator. For~both groups, the~sets of Killing vector fields are the same for isotropic and non-isotropic Petrov spaces, so in the case of group $G_3(IX)$ acting on an isotropic Petrov space, the~same local coordinate system $ \{u^a\}$ is chosen as for the non-isotropic case (Petrov~\cite{16} (f. 25.5) p. 157).
 As to the case of group $G_3(VIII)$ acting on an isotropic Petrov space, this cannot be carried out because two spaces were omitted when classifying non-isotropic Petrov spaces of type $V_3(VIII)$ (see Petrov~\cite{16} (f. 25.4) p. 157). Therefore, we accept the coordinate systems used by Petrov when classifying the spaces of type $V^{*}_4(VIII)$ (\cite{16}, formulas (25.35)--(25.37) p. 166).

\subsection{${Group \quad G_{3}(VIII)}$}

 The structural equations are of the following form:
\begin{equation}\label{8.1}
\left[{\rm X} _{1} {\rm X} _{2} \right]={\rm X} _{1}, \quad \left[{\rm X} _{2}{\rm X}_1 \right]={\rm X} _{3}, \quad \left[{\rm X} _{1} {\rm X} _{3} \right]=2{\rm X} _{2}.
\end{equation}
 As already noted, the~case   $\xi _{\left(2\right)}^{\alpha } ={\rm A}_2 p_{1} $  must be omitted, since from the structural equations it follows that   ${\rm A}_2 =1$ $\Rightarrow$ ${\rm X}_{1} $   commutes with operators  ${\rm X} _{1},  {\rm X} _{3} $,  which is impossible.
From the structural equations of \eqref{8.1}, it follows that
 $${\rm X}_{1} =\tilde{X}_{1}\exp(-u^{2}), \quad {\rm X} _{2} =\tilde{X}_{2}\exp(u^{2}), $$
where $\tilde{X}_{1} =a_{1} p_{1} +b_{1} p_{2} +r_{1} p_{3}, \tilde{X}_{3} =a_{3} p_{1} +b_{3} p_{2} +r_{3} p_{3}, \tilde{X}_{1,3} =\tilde{X}_{3,3} =0.$
Without loss of generality, one can assume $r_{1} \ne 0$. Using the admissible transformations of variables, it is possible to reduce the operators ${\rm X} _{a}$ to the following form:
\begin{equation}\label{8.3}{\rm X} _{1} =p_{3} \exp \left(-u^{2} \right), \quad{\rm X} _{2} =p_{2}, \quad {\rm X} _{3} =\left(p_{1} -2u^{3} p_{2} +\left({u^{3}}^{2} -\varepsilon \right)\right)\exp u^{2}\quad (\varepsilon = 0, \pm 1).
\end{equation}
These operators of  group $G_3( VIII )$ for the case of null Petrov spaces were found by Petrov in~\cite{16} (p. 165).
It is impossible to turn the parameter $\varepsilon$ to zero, even by coordinate transformations of the general form \eqref{2}. Since the types of solutions of the Killing equations and equations \eqref{14} depend essentially on the values of the parameter $\varepsilon$, the~relations \eqref{8.3} represent three non-equivalent sets of operators ${\rm X}_a$, each of which corresponds to a non-equivalent set of operators of the group $G_3(VIII)$, which acts in the non-isotropic Petrov space $V_4(VIII)$.
 Thus, the list of non-null homogeneous Petrov spaces of type $V_4(VIII)$ (see~\cite{16}, f. 25.5) should be supplemented.
Using the following matrices,

 $\xi _{\left(a\right)}^{\alpha } =\left(\begin{array}{ccc} {0} & {1} & {0} \\ {0} & {u^{2} } & {1} \\ {-\exp u^{3} } & {\left({u^{2}}^{2} +\varepsilon \exp 2u^{3} \right)} & {2u^{2} } \end{array}\right),$
\[\xi_{\alpha }^{\left(a\right)} =\left(\begin{array}{ccc} {\varepsilon \exp u^{3} -{u^{2}}^{2} \exp \left(-u^{3} \right)} & {2u^{2} \exp \left(-u^{3} \right)} & {-\exp \left(-u^{3} \right)} \\ {1} & {0} & {0} \\ {-u^{2} } & {1} & {0} \end{array}\right),\]
\[\xi_{\left(a\right),\gamma }^{\alpha } =\delta _{\gamma }^{2} \left(\begin{array}{ccc} {0} & {0} & {0} \\ {0} & {1} & {0} \\ {0} & {2u^{2} } & {2} \end{array}\right)+\delta _{\gamma }^{3} \left(\begin{array}{ccc} {0} & {0} & {0} \\ {0} & {0} & {0} \\ {-\exp u^{3} } & {2\varepsilon \exp 2u^{3} } & {0} \end{array}\right),\]
one can obtain the system of Equation \eqref{14} in the following form:
\begin{equation} \label{8.4}
\ell _{\left(a\right),1}^{1} =\ell _{\left(a\right)}^{3}, \quad \ell _{\left(a\right),1}^{2} =-2\varepsilon\ell _{\left(a\right)}^{3} \exp u^{3}, \quad \ell _{\left(a\right),1}^{3} =-2\ell _{\left(a\right)}^{2} \exp \left(-u^{3} \right),
\end{equation}
\begin{equation}\label{8.5}
 \ell _{\left(a\right),3}^{1} =\ell _{\left(a\right),3}^{3}= 0, \quad \ell _{\left(a\right),3}^{2} =\ell _{\left(a\right)}^{2}, \quad \ell _{\left(a\right),2}^{\alpha } =0.
 \end{equation}
From the set of Equation \eqref{8.4}, it follows that
\begin{equation}\label{8.6}
\ell _{\left(a\right)}^{1} =f^1_a(u^1), \quad \ell _{\left(a\right)}^{2} =f^2_a(u^1)\exp u^3, \quad \ell _{\left(a\right)}^{3} =f^3_a(u^1). \quad
\end{equation}
Let us substitute \eqref{8.6} to \eqref{8.5}. As a~result, we obtain the following:
\begin{equation}\label{8.7}
\dot{f}_{\left(a\right)}^1 =f_{\left(a\right)}^{3} \quad \dot{f}_{\left(a\right)}^{2} =-2\varepsilon f_{\left(a\right)}^{3} \quad \dot{f}_{\left(a\right)}^{3} =-2f_{\left(a\right)}^{2}
\end{equation}
The dot denotes the derivative of $u^{1}$.
Depending on the values of the parameter $\varepsilon $, one can obtain following solutions of the set of Equation \eqref{8.7}:

\noindent
${\bf 1}.~\varepsilon =0\Rightarrow f_{\left(a\right)}^{2} =-C_{\left(a\right)}^{3} ,  f_{\left(a\right)}^{1} =C_{a}^{3} {u^{1}}^{2} +C_{a}^{2} u^{1} +C_{a}^{1} , f_{\left(a\right)}^{3} =2C_{a}^{3} u^{1} +C_{a}^{2} \quad (C_{a}^{\alpha}=const).$
Hence the operators $Y_{a} $ can be represented as follows:
\vspace{-6pt}
\begin{equation} \label{8.8}
{\rm Y}_{1} ={\rm p}_{1} \quad Y_{2} ={\rm p}_{3} +u^{1} {\rm p}_{1} , \quad {\rm Y}_{3} ={u^{1}}^{2} {\rm p}_{1} +2u^{1} {\rm p}_{3} -{\rm p}_{2} \exp u^{3} \Rightarrow
\end{equation}
\begin{equation}\label{8,9}
\ell _{\left(a\right)}^{\alpha } = \delta^\alpha_1(\delta^1_a + {u^1}^2\delta^2_a + u^1\delta^3_a) - \exp(u^3) \delta^\alpha_2\delta^2_a + \delta^\alpha_3 (2{u^1}\delta^2_a +\delta^3_a).
\end{equation}
Matrix  $g^{\alpha \beta } = \ell _{\left(a\right)}^{\alpha } \ell _{\left(b\right)}^{\beta} \eta^{ab} $  has the following form:
\begin{equation} \label{8.10}
\left\{\begin{array}{c} {g^{11} =\left(a_{11} +2u^{1} a_{13} +{u^{1}}^{2} \left(a_{33} +2a_{12} \right)+2{u^{1}}^{3} a_{23} +a_{22} {u^{1}}^{4} \right)} \\ {g^{12} =-\exp u^{3} \left(a_{12} +a_{23} u^{1} +a_{22} {u^{1}}^{2} \right),\quad g^{22} =a_{22}\exp 2u^{3} } \\ {g^{33} =a_{33} +4u^{1} a_{23} +4{u^{1}}^{2} a_{22} ,\quad g^{23} =-\exp u^{3} \left(a_{23} +2u^{1} a_{22} \right)} \\ {g^{13} =a_{13} +u^{1} \left(a_{33} +2a_{12} \right)+3{u^{1}}^{2} a_{23} +2a_{22} {u^{1}}^{3} } \end{array}\right.
\end{equation}

\noindent
 {\bf 2}. $\varepsilon= \pm 1.$ Let us introduce the functions $st(u_{1}), ct(u^{1}) $, which, depending on the value of the parameter $\varepsilon $, have the following  form:
$$st(u)=\frac{\exp(\sqrt{\varepsilon}u)-\exp(-\sqrt{\varepsilon}u) }{2\sqrt{\varepsilon}},   ct(u)=\frac{\exp(\sqrt{\varepsilon}u)+\exp(-\sqrt{\varepsilon}u) }{2}.$$
Then, solutions of Equations \eqref{8.5}--\eqref{8.7} can be represented in the following  form:
\begin{equation}\label{8,11}
\ell _{\left(a\right)}^{\alpha } = \delta^\alpha_1(\delta^1_a + st(2{u^1})\delta^2_a + \varepsilon ct(2{u^1})\delta^3_a)+ 2\delta^\alpha_2(\delta^2_a st(2{u^1}) + \varepsilon \delta^3_a ct(2{u^1}))\exp u^3 + 2\delta^\alpha_3 (ct(2{u^1})\delta^2_a + st(2{u^1})\delta^3_a).
\end{equation}
The components of the matrix $g^{\alpha \beta } $ are as follows:
\begin{equation} \label{8.12}
\left\{\begin{array}{l} {g^{11} =\left[a_{11} +\frac{\left(a_{33} -\varepsilon a_{22} \right)}{2} \right]+ct(4u^{1}) \left(\frac{a_{33} +\varepsilon a_{22} }{2}\right)+\varepsilon a_{23} st(4u^{1}) +2\left(a_{12} st(2u^{1}) +\varepsilon a_{13} ct(2u^{1}) \right)} \\
 {g^{12} =\left[\frac{\left(a_{33} -\varepsilon a_{22} \right)}{2}+ct(4u^{1}) \left(\frac{a_{33} +\varepsilon a_{22} }{2} \right)+2\varepsilon a_{23} st(4u^{1}) +2a_{12} st(2u^{1}) +2a_{13} ct(2u^{1}) \right]\exp u^{3} } \\
  {g^{22} =2\exp 2u^{3} \left[\left(a_{33} -\varepsilon a_{22} \right)+\left(a_{33} +\varepsilon a_{22} \right)ct4u^{1} +2\varepsilon a_{23} st(4u^{1}) \right]} \\
 {g^{33} =2\left[\left(a_{22} -\varepsilon a_{33} \right)+\left(a_{22} +\varepsilon a_{33} \right)ct(4u^{1}) +2a_{23} st(4u^{1}) \right]} \\
   {g^{13} =2a_{12} ct(2u^{1}) +2a_{13} st(2u^{1}) +\left(a_{22} +\varepsilon a_{33} \right)st(4u^{1}) +2a_{23} ct(4u^{1}) } \\
 g^{23} =\left[2\left(a_{22} +\varepsilon a_{33} \right)st(4u^{1}) +4\varepsilon a_{23} ct(4u^{1}) \right]\exp u^{3}  \end{array}\right.
\end{equation}
The difference between the null space case and the non-null space case is that in the case of the space $V_3^{*}(VIII)$ the function $a_{11}$ can be converted to zero (as is actually done in formulas (25.35)--(25.37) on p. 166 in~\cite{16}). In~the case of non-null space, this cannot be~carried out.

\subsection{${Group \quad G_{3}(IX)}$}

The structural equations have the following form:
\begin{equation}\label{9.1}
\begin{array}{ccc} {\left[{\rm X} _{1} {\rm X} _{2} \right]={\rm X} _{3} ,} & {\left[{\rm X} _{2} {\rm X} _{3} \right]={\rm X} _{1} ,} & {\left[{\rm X} _{3} {\rm X} _{1} \right]={\rm X} _{2} } \end{array}.
\end{equation}
 A vector $\xi _{\left(1\right)}^{\alpha }$ will be chosen to diagonalize. Obviously, the~first variant leads to degeneracy of the set. Therefore, without restriction of generality, one can assume ${\rm X} _{1} =p_{2} $. From~the first and third equations of the structure of \eqref{9.1}, it follows that
\begin{equation} \label{9.2}
\begin{array}{cc} {{\rm X} _{2,2} ={\rm X} _{3} ,} & {{\rm X} _{3,2} =-{\rm X} _{2} \Rightarrow {\rm X}_{2,22} +{\rm X} _{2} =0} \end{array}
\end{equation}
The solution can be presented in the following  form: ${\rm X} _{2} =\tilde{X}_{2} \sin u^{2} +\tilde{X}_{3} \cos u^{2},  {\rm X}_{3} =\tilde{X}_{2}\cos u^{2} -\tilde{X}_{3} \sin u^{2} $, where the operators $\tilde{X}_{p} $ have the following  form:
\begin{equation} \label{9.3}
\widetilde{{\rm X} }_{p} =a_{p} p_{1} +b_{p} p_{2} +r_{p} p_{3}
\end{equation}
Without loss of generality, one can believe that $r_{3} =1$. Then, by admissible transformations of variables that conserve  the form of the operator ${\rm X} _{1} $, the~functions $a_{3},  b_{3} $ can be converted to zero. Thus, the~operators $\tilde{X}_{p}$ have the following form:
\begin{equation} \label{9.4}
\tilde{X}_{3} =p_{3}, \quad \tilde{X}_{2} =a_{2} p_{2} +b_{2} p_{2} +r_{2} p_{3}.
\end{equation}
From the second equation of system \eqref{9.1}, it follows that
\begin{equation} \label{9.5}
\begin{array}{ccc} {{\rm X} _{1} =p_{2} ,} & {{\rm X} _{2} =p_{3} \cos u_{2} +\frac{\sin u^{2} }{\cos u_{3} } \left(p_{1} +p_{2} \sin u^{3} \right),} & {{\rm X} _{3} ={\rm X} _{2,2} } \end{array}.
\end{equation}
This form corresponds to Petrov~\cite{16} (f. (25.6), p. 157).
Let us find the solution of the system of Equation~(14).
To do this, we use the following  matrices:
\begin{equation}\label{9.6}
 \xi _{\left(a\right)}^{\alpha } =\left(\begin{array}{lll} {\frac{\sin u^{2} }{\cos u^{3} } } & {\sin u^{2} \frac{\sin u^{3}}{\cos u^3} } & {\cos u^{2} } \\ {0} & {1} & {0} \\ {\frac{\cos u^{2} }{\cos u^{3} } } & {\cos u^{2}\frac{\sin u^{3}}{\cos u^3} } & {-\sin u^{2} } \end{array}\right), \quad \xi _{\alpha }^{\left(a\right)} =\left(\begin{array}{lll} {\sin u^{2} \cos u^{3} } & {-\sin u^{3} } & {\cos u^{2} \cos u^{3} } \\ {0} & {1} & {0} \\ {\cos u^{2} } & {0} & {-\sin u^{2} } \end{array}\right),
\end{equation}
\begin{equation*}\label{9.8}
\xi _{\beta }^{\left(b\right)} \xi _{\left(b\right),\gamma }^{\alpha } =\delta _{\gamma }^{2} \left(\begin{array}{lll} {0} & {0} & {-\cos u^{3}} \\ {0} & {0} & {0} \\ {\frac{1}{\cos u^{3} } } & {\frac{\sin u^{3}}{\cos u^3}} & {0} \end{array}\right)+\delta _{\gamma }^{3} \left(\begin{array}{lll} {\frac{\sin u^{3}}{\cos u^3} } & {\frac{1}{\cos u^{3} } } & {0} \\ {0} & {0} & {0} \\ {0} & {0} & {0} \end{array}\right)
\end{equation*}
The system of Equation~(14) will take the following  form:
\begin{equation} \label{9.9}
\left\{\begin{array}{lll} {\ell _{\left(a\right),3}^{1} =\frac{\ell _{\left(a\right)}^{2} }{\cos u^{3} } } & {\ell _{\left(a\right),3}^{2} =\ell _{\left(a\right)}^{2} \frac{\sin u^{3}}{\cos u^3} } & {\ell _{\left(a\right),3}^{3} =0 } \\ {\ell _{\left(a\right),1}^{1} =\ell _{\left(a\right)}^{3} \frac{\sin u^{3}}{\cos u^3} } & {\ell _{\left(a\right),1}^{2} =\frac{\ell _{\left(a\right)}^{3} }{\cos u^{3} } } & {\ell _{\left(a\right),1}^{3} =-\cos u^{3}\ell _{\left(a\right)}^{2} } \end{array}\right.
\end{equation}
The system \eqref{9.9} has a solution:
$$
{\rm Y}_{1} ={\rm p}_{1}, \quad Y_{2} =\frac{\sin u^3}{\cos u^3}({\rm p}_{2} +{\rm p}_{1}\sin u^3) +{\rm p}_{3}\cos{u^1}, \quad {\rm Y}_{3} ={\rm Y}_{2,1} \Rightarrow
$$
\begin{equation*} \label{9.10}
\ell _{\left(a\right)}^{\alpha } = \delta^\alpha_1(\delta^1_a + \frac{\sin u^3}{\cos u^3}(\delta^2_a \sin u^1 + \delta^3_a\cos{u^1}))+ \frac{\delta^\alpha_2 }{\cos u^3}(\delta^2_a \sin u^1 + \delta^3_a\cos{u^1})
+\end{equation*}
$$+\delta^\alpha_3(\delta^2_a \cos{u^1} -\delta^3_a\sin u^1).$$
The components of the matrix $g^{\alpha \beta } $ are as follows:
\begin{equation} \label{8.12}
\left\{\begin{array}{l} {g^{11} =a_{11} -2\frac{\cos u^3}{\sin u^3}(a_{12}\sin u^1+ a_{13}\cos
u^1) +  ({\frac{\cos u^3}{\sin u^3}})^2 (\frac{a_{33} + a_{22} }{2} + (\frac{a_{33} - a_{22} }{2})\cos 2u^1+a_{23} \sin2u^{1})}

\\ {g^{12} =\frac{1}{\sin u^3}(a_{12}\sin u^1+ a_{13}\cos
u^1 - {\frac{\cos u^3}{\sin u^3}} (\frac{a_{33} + a_{22} }{2} + \frac{a_{33} - a_{22} }{2}\cos 2u^1+a_{23} \sin2u^{1}))}

\\ {g^{22} =\frac{1}{\sin^2 u^3}(\frac{a_{33} + a_{22} }{2} + \frac{a_{33} - a_{22} }{2}\cos 2u^1+a_{23} \sin2u^{1})} \\
 {g^{33} =\frac{1}{\sin u^3}(\frac{a_{33} + a_{22} }{2} - \frac{a_{33} - a_{22} }{2}\cos 2u^1-a_{23} \sin2u^{1})}
  \\
   {g^{13} =a_{12}\cos u^1 - a_{13}\sin u^1 - {\frac{\cos u^3}{\sin u^3}} (\frac{a_{22} - a_{33}}{2} \sin 2u^1+a_{23} \cos2u^{1})} \\
 g^{23} =\frac{a_{33} + a_{22} }{2} + \frac{a_{22}-a_{33}}{2}\cos 2u^1-a_{23} \sin2u^{1}  \end{array}\right.
\end{equation}

\section{{ List  of  Obtained~Results}}\label{sec6}

In this section, all non-equivalent sets of operators of groups $G_3(N)$ acting simply transitively in homogeneous Petrov spaces $V_4(N)$ and $V^{*}_4(N)$ are given. The~generators of infinitesimal transformations are given by Killing vector fields:
${\rm X}_a = \xi^i_a {\rm p}_i$. The~generators of non-infinitesimal transformations are given by the vector fields of the reper: ${\rm Y}_a = \ell^i_a {\rm p}_i$.
All sets of operators ${\rm X}_a$, except~for those acting in Petrov spaces $V_4(VIII)$, are equivalent to those in Petrov's book~\cite{16} (formulas (25.1)--(25.6) and (25.24)--(25.38)). In~contrast to Petrov's book, all solutions for spaces $V^{*}_4(VIII)$ are given in the canonical semi-geodesic coordinate system \eqref{6}. The~contravariant components of the metric tensor of the spaces are given in Sections~\ref{sec5} and \ref{sec6}.
All sets of operators of each group that are non-equivalent with respect to the admissible coordinate transformations of the form \eqref{8} are equivalent with respect to admissible coordinate transformations of the form \eqref{2}. The~exception is the group $G_3(N)$.
 Each of the three sets of operators of the group $G_3(VIII)$ is non-equivalent to the other two sets with respect to both coordinate transformations \eqref{2} and \eqref{8}.

\noindent
1. Group $G_3(II)$

\noindent
First version
\begin{equation} \label{7.1}
{\rm X} _{1} ={\rm p}_{1}, \quad {\rm X} _{2} ={\rm p}_{2}, \quad {\rm X} _{3} = u^{2} {\rm p}_{1} +{\rm p}_{3}.
\end{equation}
\begin{equation} \label{7.2}
{\rm Y } _{1} ={\rm p}_{1}, \quad {\rm Y} _{2} = u^{3} {\rm p}_{1} +{\rm p}_{2},\quad{\rm Y} _{3} ={\rm p}_{3}.
\end{equation}

\noindent
Second version
\begin{equation} \label{7.3}
{\rm X} _{1} ={\rm p}_{2}, \quad {\rm X} _{2} ={\rm p}_{3}, \quad  {\rm X} _{3}= {\rm p}_{1} + u^{3} {\rm p}_{2},\quad \tilde{{\rm X}} _{3} = {\rm X} _{3}+{\ell_{0}}(u^0) {\rm p}_{3}.
\end{equation}
\begin{equation} \label{7.4}
{\rm Y} _{1} ={\rm p}_{1}, \quad {\rm Y} _{2} ={\rm p}_{2}, \quad {\rm Y} _{3}= {\rm p}_{1} + u^{3} {\rm p}_{2}.
\end{equation}

\noindent
2. Group $G_3(III)$

\noindent
First version
\begin{equation} \label{7.5}
{\rm X} _{1} =\rm p_{1}, \quad {\rm X} _{2} =\rm p_{2}, \quad {\rm X} _{3} =\rm p_{3} +u^{2}\rm p_{2},
\end{equation}
\begin{equation} \label{7.6}
{\rm Y} _{1} ={\rm p}_{1}, \quad {\rm Y} _{2} ={\rm p}_{2}\exp u^3, \quad {\rm Y} _{3} = {\rm p} _{3}.
\end{equation}

\noindent
Second version

\begin{equation} \label{7.7}
{\rm X} _{1} ={\rm p}_{2}, \quad {\rm X} _{2} ={\rm p}_{3}, \quad  {\rm X}_{3} ={\rm p} _{1} +u^{3}{\rm p}_{3}, \quad\tilde{{\rm X}} _{3} = {\rm X} _{3}+{\ell_{0}}(u^0) {\rm p}_{2}.
\end{equation}
\begin{equation} \label{7.8}
{\rm Y} _{1} ={\rm p}_{1}, \quad {\rm Y} _{2} ={\rm p}_{2}, \quad {\rm Y} _{3} = {\rm p}_{3}\exp u^1.
\end{equation}

\noindent
Third version
\begin{equation} \label{7.9}
{\rm X}_{1}= {\rm p}_{2}, \quad {\rm X} _{2} ={\rm p}_{1}\exp{-u^{3}}, \quad {\rm X} _{3} ={\rm p}_{3}.
\end{equation}
\begin{equation} \label{7.10}
{\rm Y} _{1} ={\rm p}_{1}, \quad {\rm Y} _{2} = {\rm p}_{2},\quad {\rm Y} _{3} ={\rm p}_{3} - u^1{\rm p}_{1}.
\end{equation}

\noindent 3. Group $G_3(IV)$

\noindent
First version
\begin{equation} \label{7.11}
{\rm X} _{1} ={\rm p} _{1} {\rm exp}\left(-u^{3} \right), \quad {\rm X} _{2} ={\rm p} _{2}, \quad{\rm X} _{3} ={\rm p} _{3} +u^{2} \left({\rm p} _{1} {\rm exp}\left(-u_{3} \right)+{\rm p} _{2} \right)
\end{equation}
\begin{equation} \label{7.12}
{\rm Y} _{1} ={\rm p} _{1}, \quad {\rm Y} _{2} ={\rm p} _{1} u^{3}  + {\rm p}_{2}\exp u_{3}, \quad
{\rm Y} _{3} ={\rm p} _{3} -{\rm p} _{1} u^{3},
\end{equation}

\noindent
Second version
\begin{equation} \label{7.13}
{\rm X} _{1} =p_{2} ,\quad {\rm X} _{2} =p_{3},\quad {\rm X}_{3} =p_{1} +(u^{2} + u^{3} )p_{2}+u^{3}p_{3}.
\end{equation}
\begin{equation} \label{7.14}
{\rm Y} _{1} ={\rm p} _{1}, \quad {\rm Y} _{2} = {\rm p}_{2}\exp u^{1}, \quad
{\rm Y} _{3} =({\rm p} _{3}+{\rm p} _{1} u^{1})\exp u^{1},
\end{equation}

\noindent
Third version
\begin{equation}\label{7.15}
{\rm X}_{1} = {\rm p}_{2}, \quad {\rm X}_{2} = {\rm p}_{1}\exp{(-u^3)} - {\rm p}_{2} u^3, \quad {\rm X}_{3} = {\rm p}_{3} +  {\rm p}_{2} u^2.
\end{equation}
\begin{equation} \label{7.16}
{\rm Y} _{1} ={\rm p} _{2}\exp u^{3}, \quad {\rm Y} _{2} = {\rm p}_{1}, \quad
{\rm Y} _{3} ={\rm p} _{3}+u^1({\rm p} _{1} +{\rm p} _{2}\exp u^{3}),
\end{equation}

\noindent
{4. Group $G_3(V)$

\noindent
First version
\begin{equation} \label{7.17}{\rm X} _{1} ={\rm p} _{2}, \quad {\rm X} _{2} ={\rm p} _{3}, \quad  {\rm X} _{3} ={\rm p} _{1} +u^{2} {\rm p} _{2} +u^{3} {\rm p} _{3}
\end{equation}
\begin{equation} \label{7.18}
{\rm Y} _{1} ={\rm p}_{1}, \quad {\rm Y} _{2} = {\rm p}_{2}\exp u^{1},\quad {\rm Y} _{2} = {\rm p}_{2}\exp u^{1}. \end{equation}

\noindent
Second version
\begin{equation}\label{7.19}
{\rm X} _{1} ={\rm p }_{2}, \quad{\rm X} _{2} ={\rm p } _{1} {\exp}u_{3}, \quad {\rm X} _{3} =-{\rm p }_{3} + u^2{\rm p }_{2}, \quad\tilde{{\rm X}} _{2} ={\rm X} _{2} +\ell_{0} {\rm p }_{2}.
\end{equation}
\begin{equation}\label{7.20}
{\rm Y}_{1} ={\rm p}_{1}, \quad {\rm Y} _{2} = {\rm p}_{2}\exp u^{3},\quad {\rm Y}_{2} = {\rm p}_{2} + u^1 {\rm p}_{1}.
\end{equation}

\noindent
5. Group $G_3(VI)$

\noindent
First version
\begin{equation} \label{7.21}
{\rm X} _{1} =p_{1} {\rm exp}\left(-u^{3} \right), \quad {\rm X} _{2} ={\rm p}_{2}, \quad {\rm X}_{3} ={\rm p}_{3} +qu^{2} {\rm p}_2.
\end{equation}
\begin{equation} \label{7.22}
Y_{1} =p_{1}, \quad Y_{2} =p_{2} {\exp}(qu^{3}), \quad Y_{3} =p_{3} -u^{1} p_{1}.
\end{equation}

\noindent
Second version
\begin{equation}\label{7.23}
{\rm X} _{1} =p_{2}, \quad {\rm X} _{2} =p_{3}, \quad
{\rm X} _{3} =p_{1} +u^{2} p_{2} +qu^{3} p_{3}.
\end{equation}
\begin{equation} \label{7.24}
Y_{1} =p_{1}, \quad Y_{2} =p_{2} {\exp}u^{1}, \quad Y_{3} =p_{3}{\exp}(qu^{1}).
\end{equation}

\noindent
Third version
\begin{equation} \label{7.25}
{\rm X} _{1} ={\rm p}_{2}, \quad {\rm X} _{2} ={\rm}p_{1} \exp(-qu^{3}) +{\rm p}_{2} \exp((1-q)u^{3}), \quad {\rm X} _{3} ={\rm p}_{3} +u^{2} {\rm p}_2.
\end{equation}
\begin{equation} \label{7.26}
{\rm Y}_{1} =p_{1}, \quad {\rm Y}_{2} ={\rm p}_{2} {\exp}u^{3}, \quad {\rm Y}_{3} =p_{3}+u^1({\rm p}_{2}(1-q){\exp}u^{3}-q{\rm p}_{1}).
\end{equation}

\noindent
Group $G_3(VII)$

\noindent
First version
\begin{equation}\label{7.27}
{\rm X} _{1} =p_{2}, \quad {\rm X} _{2} =p_{3}, \quad {\rm X} _{3} =p_{1} +u^{3} (2\cos{\alpha} p_{3} -p_{2})+u^{2} p_{3}.
\end{equation}
\vspace{-24pt}
\begin{equation}\label{7.28}
Y_{1} =p_{1},\quad Y_{2} =\left(p_{2} \sin p-p_{3} \sin \left(p+\alpha \right)\right)\exp q,\quad Y_{3} =\left(p_{2} \cos{p}-p_{3} \cos\left(p+\alpha \right)\right)\exp q
\end{equation}
$(p=u^{3} \sin \alpha ,\quad q=-u^{3} \cos\alpha ).$

\noindent
Second version
\vspace{-6pt}
\begin{equation}\label{7.29}
 {\rm X} _{1} ={\rm p}_{2} ,\quad {\rm X} _{3} ={\rm p}_{3} +u^{2} {\rm X} _{2} ,\quad {\rm X} _{2} ={\rm p}_{1}\frac{\exp - q}{\cos p} +{\rm p}_{2} \frac{\cos \left(p-\alpha \right)}{\cos p} .
 \end{equation}
\begin{equation} \label{7.30}
Y_{1} =p_{2} ,\quad Y_{2} =p_{1}\sin \alpha\frac{\sin p}{\cos p} +p_{2}\sin^2(\alpha)\frac{\exp q}{\cos p}, \quad  Y_{3} =p_{3} +u^{1} (Y_{2}-p_{1}\cos \alpha ).
\end{equation}

\noindent
Group $G_3(VIII)$

\begin{equation}\label{7.31}{\rm X} _{1} =p_{3} \exp \left(-u^{2} \right), \quad{\rm X} _{2} =p_{2}, \quad {\rm X} _{3} =\left(p_{1} -2u^{3} p_{2} +\left({u^{3}}^{2} -\varepsilon \right)\right)\exp u^{2}\quad (\varepsilon = 0, \pm 1).
\end{equation}

\noindent
First version  $\varepsilon =0$
\begin{equation} \label{7.32}
{\rm Y}_{1} ={\rm p}_{1} \quad Y_{2} ={\rm p}_{3} +u^{1} {\rm p}_{1} , \quad {\rm Y}_{3} ={u^{1}}^{2} {\rm p}_{1} +2u^{1} {\rm p}_{3} -{\rm p}_{2} \exp u^{3}.
\end{equation}

\noindent
Second version  $\varepsilon^2 = 1$
\begin{equation}\label{7.33}
{\rm Y}_{1} =p_{1}, \quad {\rm Y}_{2} ={\rm p}_{1}st(2u^1) + 2{\exp}u^{3}st(2u^1){\rm p}_{2} + 2ct(2u^1){\rm p}_{3}, \quad {\rm Y}_{3} = \varepsilon{\rm Y}_{2,1},
\end{equation}
where the following functions are introduced:
$$st(u)=\frac{\exp(\sqrt{\varepsilon}u)-\exp(-\sqrt{\varepsilon}u) }{2\sqrt{\varepsilon}},  \quad ct(u)=\frac{\exp(\sqrt{\varepsilon}u)+\exp(-\sqrt{\varepsilon}u) }{2}.$$

\noindent
Group $G_3(IX)$
\begin{equation} \label{7.34}
\begin{array}{ccc} {{\rm X} _{1} =p_{2} ,} & {{\rm X} _{2} =p_{3} \cos u^{2} +\frac{\sin u^{2} }{\cos u^{3} } \left(p_{1} +p_{2} \sin u^{3} \right),} & {{\rm X} _{3} ={\rm X} _{2,2} } \end{array}.
\end{equation}
$$
{\rm Y}_{1} ={\rm p}_{1}, \quad Y_{2} =\frac{\sin u^3}{\cos u^3}({\rm p}_{2} +{\rm p}_{1}\sin u^3) +{\rm p}_{3}\cos{u^1}, \quad {\rm Y}_{3} ={\rm Y}_{2,1}.
$$

\section{Conclusions} 
}

In the theory of gravitation, a special place is occupied by Riemannian manifolds, on the~spacelike hypersurfaces of which three-parameter groups of motions act simply transitively. These manifolds, which are usually called homogeneous spaces of type $N$ by Bianchi, generalize the homogeneous isotropic model of the Universe (see~\cite{36a}). They are models of a homogeneous but non-isotropic Universe and may be of interest, for~example, in~the early stages of the Universe's life. Obviously, the~traditional models of the Universe are special cases of these spaces. Homogeneous Petrov spaces, the~classification of which is completed in the present paper, generalize these homogeneous spaces in the following way. First, in~homogeneous Petrov spaces, non-null hypersurfaces of transitivity  can be time-like. Second, these hypersurfaces can be isotropic.
Homogeneous Petrov spaces with null hypersurfaces of transitivity can serve as models of plane-wave metrics (examples of the study of such spaces can be found in~\cite{21,22,23}).  The~geometry of homogeneous Petrov spaces in both cases is determined by the geometry of transitivity spaces $V_3(N)$ (in the null case under the condition that the Killing vector fields are independent of $u^0$).

Note that homogeneous Petrov spaces belong to the type of spaces admitting three Killing fields (vector or tensor). Besides~them, such a feature is also possessed by Stackel spaces (see, for~example,~\cite{37d} and the bibliography therein). For~spaces of this type there are methods of exact integration of the equations of motion of test bodies. Gravitational equations in these spaces can be reduced to systems of ordinary differential equations. The~classification carried out in this paper facilitates the transition to these systems. Thus, the~classification of all Riemannian spaces with Lorentz signature admitting a triple of Killing fields is completed (see also~\cite{33b,35c}).

\quad

This work is supported by the Russian Science Foundation, Project No. N 23-21-00275.


%



\end{document}